\documentclass[12pt,preprint]{aastex}

\shorttitle{M31 globulars and the fundamental plane}
\shortauthors{Barmby et al.}

\begin{document}

\title{Structural parameters for globular clusters in M31 and generalizations 
for the fundamental plane%
\footnote{
Based on observations made with the NASA/ESA Hubble Space Telescope, obtained at the Space 
Telescope Science Institute, which is operated by the Association of Universities for Research 
in Astronomy, Inc., under NASA contract NAS 5-26555. These observations are associated with 
programs \#8664, \#9719, \# 9767 and \#10260.}
}

\author{Pauline Barmby}
\affil{Harvard-Smithsonian Center for Astrophysics, 60 Garden St., Mailstop 65, Cambridge, MA 02138}
\author{Dean E. McLaughlin}
\affil{Department of Physics and Astronomy, University of Leicester, University Road, Leicester LE1 7RH, UK}
\author{William E. Harris}
\affil{Department of Physics and Astronomy, McMaster University, Hamilton, ON L8S 4M1, Canada}
\author{Gretchen L.H. Harris}
\affil{Department of Physics and Astronomy, University of Waterloo, Waterloo, ON N2L 3G1, Canada}
\and
\author{Duncan A. Forbes}
\affil{Centre for Astrophysics and Supercomputing, Swinburne University, Hawthorn, VIC 3122, Australia}

\begin{abstract}
The structures of globular clusters (GCs) reflect their dynamical
states and past histories. High-resolution imaging allows the exploration of 
morphologies of clusters in other galaxies.
Surface brightness profiles from new {\it Hubble Space Telescope} observations
of 34 globular clusters in M31 are presented, together with fits of several different
structural models to each cluster. 
M31 clusters appear to be adequately fit by standard King models, and do not
obviously require alternate descriptions with relatively stronger halos, such
as are needed to fit many GCs in other nearby galaxies.
The derived structural parameters are combined with corrected versions of
those measured in an earlier survey to construct a comprehensive catalog of
structural and dynamical parameters for M31 GCs with a sample size similar
to that for the Milky Way.
Clusters in M31, the Milky Way, Magellanic Clouds, Fornax dwarf
spheroidal and NGC~5128 define a very tight fundamental plane with 
identical slopes.
The combined evidence for these widely different galaxies strongly reinforces the view that
old globular clusters have near-universal structural properties regardless of host
environment.
\end{abstract}

\keywords{galaxies: star clusters --- galaxies: individual (M31) --- globular clusters: general}

\section{Introduction}

Globular clusters (GCs) have long been recognized as unique dynamical laboratories
and unique markers of galaxy history. The integrated properties of GCs, 
such as age and metallicity, are believed to reflect conditions in the early
stages of galaxy formation \citep{bs06}. The spatial structure and kinematics
of GCs reflect both their formation conditions and dynamical evolution within 
the tidal fields of their host galaxies.

Structures of GCs as described by surface brightness or velocity profiles 
can be fit by a number of different models. These models can be used to 
compute numerous parameters describing various aspects of cluster structure,
but four independent parameters suffice to completely define a single-mass model. It is helpful
to think of cluster structures as being defined by spatial scale, spatial concentration, 
total luminosity, and a relation between luminosity and mass, but
many different parameter choices are possible. \citet{dj95} and \citet{mcl00} used formally
different, but entirely equivalent, sets of structural parameters to show that 
globular clusters in the Milky Way do not occupy the full four-dimensional
parameter space but instead define a remarkably narrow `fundamental plane'.
The same appears to be true for clusters in M31 \citep{djo97,dg97,bhh02}, in NGC~5128 \citep{hhhm02}, 
and in M33 \citep{m33_hires}. This evidence indicates that the formation and evolution
processes for normal globular clusters were similar in a wide variety of environments \citep{har03}.

The globular cluster fundamental plane is similar in character, but different
in detail, to that of elliptical galaxies \citep[for an introduction, see][]{bur97}.
There have been several suggestions that at least some of the most massive GCs in the Milky Way
and M31 are in fact the nuclei of destroyed dwarf galaxies \citep[e.g.][]{tsc03,hr00},
or that they are related to ultra-compact dwarf galaxies \citep{has05}. Both dwarf galaxy 
nuclei and UCDs occupy somewhat different regions of the fundamental plane than do GCs,
but structural properties have been measured for only a few objects
in this transition mass region between globulars and galaxies.
Combining samples of clusters from a number of nearby galaxies---
where structures can be accurately measured with high-resolution imaging from the
{\it Hubble Space Telescope} (HST)--- is a way to improve the sample statistics
particularly for the upper end of the GC mass range.

To further understand the globular cluster fundamental plane, we have carried out 
a survey of massive globulars in M31 and
NGC~5128 using the HST Advanced Camera for Surveys \citep[ACS;][]{acs}.
Other papers from this program discuss the NGC~5128 cluster sample \citep{harris06},
the techniques for measurement of those clusters' structural parameters \citep{dean06},
and the comparison between massive globulars, nuclear star clusters, and `dwarf-globular
transition objects' \citep{dean07}.
This paper presents the new M31 observations and structural parameters,
compiles a publicly-available catalog of measurements of structural parameters for
the largest-ever sample of M31 clusters, and compares the resulting `fundamental
planes' of globular cluster parameters in different galaxies.
Throughout this paper we assume a distance to M31 of 780 kpc (1~pc  
subtends 0\farcs26)
and use the cluster nomenclature of the Revised Bologna Catalogue \citep{gal06}. 

\section{Observational material}
\label{sec:obsdat}

We have used material from several different HST programs.
Program GO-10260 (PI W. Harris) obtained ACS Wide Field Channel (WFC) images 
centered on six clusters in M31
and twelve clusters in NGC~5128, selected to be those having $M_V<-10.5$ and (for M31) no
previous HST observations. Exposure times were 2370~s (one orbit split into three dither positions)
in both the F606W and F814W filters.
Most of the GO-10260 M31 observations were made in the period 2004 September 29--2004 October 01.
Program GO-9719 (PI T. Bridges) obtained ACS High Resolution Channel (HRC) images of 
three M31 clusters using
exposure times of 2020~s in F606W and 2680~s in F814W in 2003 August and September.
Program GO-8664 (PI W. Harris) targeted 24 M31 clusters with a range of properties
\citep[as well as NGC~5128 clusters, discussed in][]{hhhm02}
for imaging with STIS in snapshot mode; exposure times were 480~s in the unfiltered
50CCD/CL imaging mode. These data were obtained in fall 2000 and winter 2001.
Of the 24 STIS targets, one (B329) is clearly a galaxy,
and two more (B391 and B340) had faulty coordinates and were not on the images.
Cluster B396 was bisected by the edge of the STIS image but the data 
were still usable.

The cluster G001 (Mayall II) is one of the most massive in M31 and has been
the subject of numerous recent investigations. It is often likened to the Milky Way's
own massive cluster $\omega$~Cen; both clusters have been suggested as possible
remnant dwarf galaxy cores \citep{mey01}. There has been some discussion about
whether G001 contains a central black hole \citep{grh02,baum03,grh05} and
also some widely differing measurements of its structural parameters
\citep[see discussion in][]{bhh02}. \citet{grh05} used new HST/ACS observations
of G001 (obtained as part of GO-9767) to produce a surface brightness profile
as part of an effort to constrain the mass of any putative central black hole, 
but those authors did not give values for the cluster's structural parameters. 
\citet{ma07} fit \citet{kin66} models to the ACS data. Because of the importance
of G001, we decided to re-analyze 
the \citet{grh05} observations, which consist of six exposures totaling 41 minutes of integration
time with the ACS HRC in the F555W filter.

\begin{figure}
\plotone{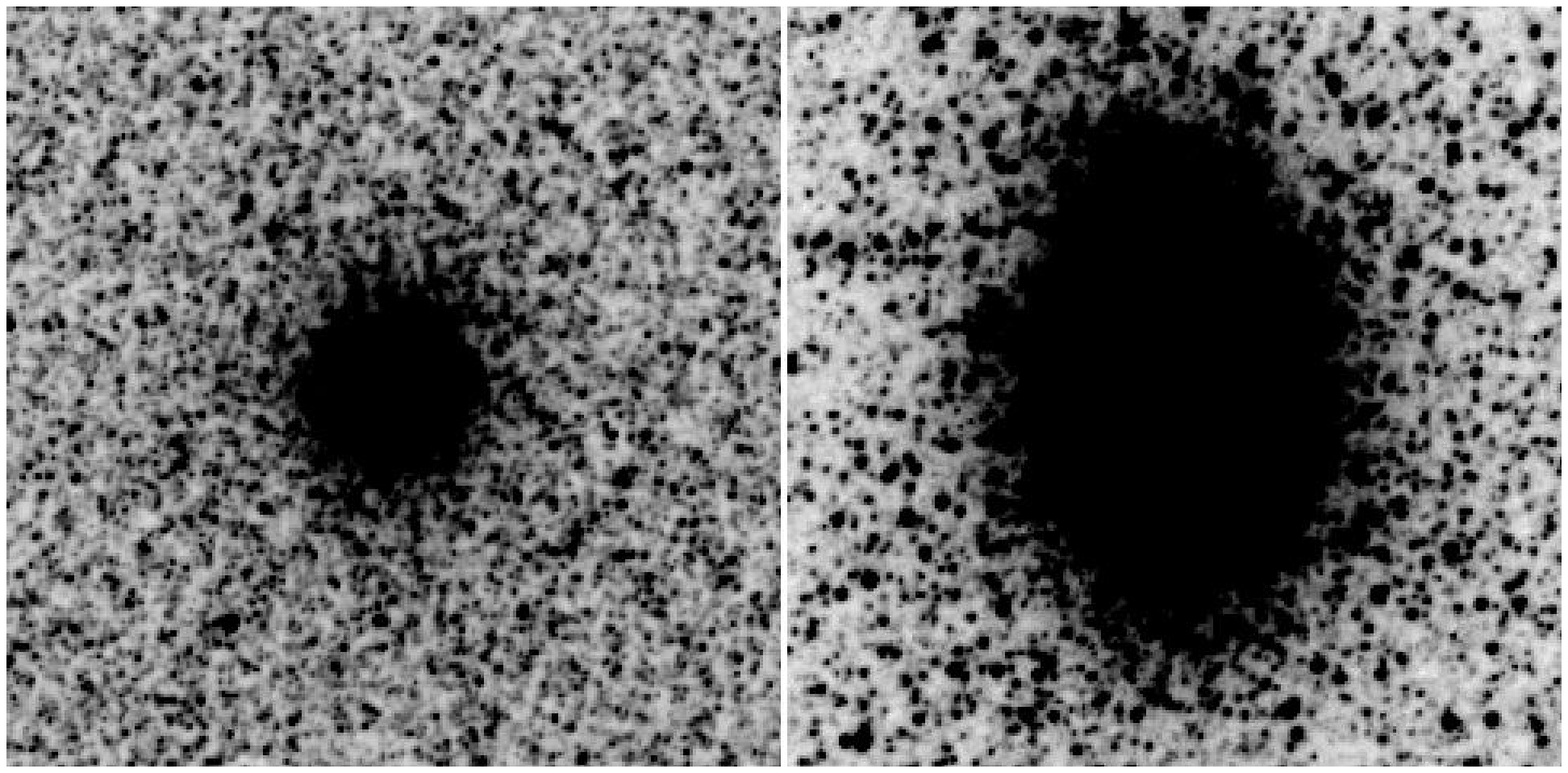}
\caption{
ACS/WFC F606W images of two M31 clusters. Left: B147,
\citep[classified as a star by][based on a velocity dispersion measurement]{dg97}. 
Right: the very elliptical ($\epsilon=0.28$) cluster B088.
Both images have north up and east to the left and are 12\arcsec\ square.
\label{fig_pix}}
\end{figure}

The HST images offered the possibility of detecting new clusters or confirming
additional cluster candidates other than the targets.
All the M31 images were searched for additional clusters and inspected at the positions of cataloged cluster 
candidates. A number of additional cluster candidates appeared in the
ACS images: B041, B056D, B061, B081D, B088D, B090, B102, B147, B162 and M027.
We also (independently) re-discovered the new cluster B515 reported by \citet{gal06}.
All of these candidates appear to be clusters except for B102, which
is apparently a star.\footnote{\citet{gal06} independently came to the same conclusions about the 
nature of B056D, B102, and B162.}
Interestingly, B147, which had been classified as a star based on velocity 
dispersion measurements by \citet{dg97}, is clearly a resolved cluster
(see Figure~\ref{fig_pix}). This object is also associated with an X--ray source
\citep{tp04}.
Figure~\ref{fig_pix} shows images of B147 and the highly elliptical cluster B088
demonstrating the power of HST imaging for resolving M31 globular clusters.

To generate the largest possible sample of M31 globulars for fundamental plane
analysis, we also include most of the clusters with structural parameters measured
from images taken with HST's Wide Field and Planetary Camera 2 (WPFC2), reported
in \citet{bhh02}. The clusters from \citet{bhh02} included in
the present sample include 59 objects with {\it HST/WFPC2} photometry in the
$V$-band and good measurements of $r_0$ and $c$.
The central surface brightness values for these clusters have had an important error corrected.
As also discussed in \citet{dean06}, the software used to
fit structural models in \citet{bhh02} and also \citet{hhhm02} did not return
a model central surface brightness value; instead both papers reported 
$\mu_{V,0}$ as the surface brightness of the central pixel of
the PSF-convolved model image. As such, the values reported in 
\citet{bhh02} are systematically too faint. We re-computed 
$\mu_{V,0}$ from the \citet{kin66} $c$ and $r_0$ values reported in
\citet{bhh02} and the integrated $V$ magnitudes reported in \citet{b00} or \citet{bh01}.
The median surface brightness change is $\Delta \mu_{V,0}=-0.33$~mag,
with a large scatter. The maximum change is about 1.5~mag, and 
there were several clusters, both bright and faint, whose central
surface brightnesses were nearly unchanged in the re-calculation.
Using the model-fitting software of both \citet{dean06} and \citet{bhh02} 
on a small number of M31 clusters showed that both recovered the
the same $r_0$ and $c$ values to within the uncertainties, so the two sets of 
measurements should be consistent.

To summarize, the new observations include 15 clusters observed using the ACS in either
the Wide Field or High-Resolution Channel, in both F606W and F814W filters; 
and 19 clusters observed in the unfiltered mode using STIS.
The observations of cluster G001 with ACS/HRC in the F555W filter are not new
but are re-analyzed here.
There are therefore 50 surface-brightness profiles for 34 distinct objects 
(one of the STIS clusters, B023, is also in the ACS 
sample and has thus been observed three times).
Adding the 59 objects from \citet{bhh02}, the sample of M31 clusters
totals 93 objects, with some later omitted from the fundamental plane
analyses (see \S~\ref{subsec:transmet}). Estimates of the total size of
the M31 GCS vary: one recent value is $N_{gc}=460\pm70$ \citep{bh01}.
Thus the HST sample comprises only a fraction of the total M31 population, but
it is comparable in size to the full sample of non-core-collapsed
Milky Way clusters considered by \citet{mcl00}, and about 10\% larger than
the sample of 85 Milky Way clusters analyzed by \citet{mv05}.
We have not attempted to generate a complete sample of clusters observed 
with HST by searching the Archive for clusters serendipitously
observed in data taken after the \citet{bh01} sample was compiled;
however we expect that the present sample should be representative of M31 GCs
particularly at the high-luminosity end. The exception is low-surface-brightness 
clusters for which our present analysis method is unsuitable: these include
both faint clusters, some discussed in the following section, and also the
`extended luminous' clusters discovered by \citet{huxor05}
(see also \S\ref{subsec:fpcor}).

\section{Data analysis}

The data analysis procedures used for these data are described in detail by
\citet{dean06}. Here the procedures are briefly summarized, and 
some details specific to the M31 observations noted.

\subsection{Surface Brightness Profiles}
\label{subsec:SBprofs}

The STScI-pipeline output drizzled images were used to
measure cluster shapes with ELLIPSE in IRAF. 
Cluster center positions were fixed at values derived by centroiding and elliptical
isophotes were fit to the data, with no sigma-clipping. 
A first pass of ELLIPSE was run in the usual way, with
ellipticity and position angle allowed to vary with the isophote semi-major axis.
Luminosity-weighted averages of ellipticity and position angle were
determined from the results. In the second pass of ELLIPSE, 
surface brightness profiles on fixed, zero-ellipticity isophotes 
were measured (this was required because we chose to fit circular models
for both the intrinsic cluster structure and the PSF).
For several low-density clusters (B290 and B423 observed with STIS;
B081D, B088D, B515, and M027 observed with ACS), 
a sensible, monotonically-decreasing profile could not be obtained:
no models were fit to these clusters.
Table~\ref{tab:m31meas} reports the average ellipticities, position angles, 
and aperture magnitudes measured from the ACS and STIS images, and tabulates some additional
reference data (metallicity, reddening, galactocentric distance) for all of the M31 clusters.
For 11 of the clusters, there are no metallicity estimates in
the literature, so we have simply assigned ${\rm [Fe/H]}=-1.2\pm0.6$ to all of
them (the mean and standard deviation of the metallicity
distribution of the M31 GC system; \citealt{b00}).

The raw output from ELLIPSE is in units of ${\rm cts/s}$ per pixel,
which is converted to ${\rm cts/s}$ per square arcsecond by multiplying by
$400=(1\,{\rm px}/0\farcs05)^2$ for WFC, $1600=(1\,{\rm px}/0\farcs025)^2$ for HRC,
or $387.5=(1\,{\rm px}/0\farcs0508)^2$ for STIS. 
To transform the ACS counts to surface brightness calibrated on the
VEGAMAG system (from the ACS Handbook)
\begin{equation}
\mu/{\rm mag\ arcsec^{-2}} = z - 2.5\,\log ({\rm cts/s}/\sq\arcsec)
\end{equation}
where the zeropoints are $z=(25.255,26.398, 25.893, 25.501, 24.849)$ for HRC/F555W, WFC/F606W,
HRC/F606W, WFC/F814W, and HRC/F814W respectively.
The STIS data were taken in unfiltered (``clear'' or CL)
mode, for which \citet{hhhm02} derive
\begin{equation}
  \mu_V/{\rm mag\ arcsec^{-2}} =
        (26.29\pm0.05) -  2.5\,\log({\rm cts/s}/\sq\arcsec)
\label{eq:stisVzp}
\end{equation}
in the standard $V$ bandpass.

We allow for the occasional over-subtraction of sky in the
automatic reduction pipeline, and hence ``negative'' counts in some pixels, by
working in terms of linear intensity instead of surface brightness in
magnitudes. Given\footnote{\url{http://www.ucolick.org/~cnaw/sun.html}} 
$M_{\odot,F606W}=4.64$, $M_{\odot,F814W}=4.14$,
and $M_{\odot,V/F555W}=4.83$, the calibration above yields
\begin{equation}
  \begin{array}{lcll}
   I_{F555W}/L_\odot\ {\rm pc}^{-2} & \simeq &
         2.8766 \times ({\rm cts/s}/\sq\arcsec) & \quad {\rm [HRC]} \\ 
   I_{F606W}/L_\odot\ {\rm pc}^{-2} & \simeq &
         0.8427 \times ({\rm cts/s}/\sq\arcsec) & \quad {\rm [WFC]} \\
   I_{F606W}/L_\odot\ {\rm pc}^{-2} & \simeq &
         1.3418 \times ({\rm cts/s}/\sq\arcsec) & \quad {\rm [HRC]} \\
   I_{F814W}/L_\odot\ {\rm pc}^{-2} & \simeq &
         1.2147 \times ({\rm cts/s}/\sq\arcsec) & \quad {\rm [WFC]} \\
   I_{F814W}/L_\odot\ {\rm pc}^{-2} & \simeq &
         2.2145 \times ({\rm cts/s}/\sq\arcsec) & \quad {\rm [HRC]} \\
   I_{V}/L_\odot\ {\rm pc}^{-2} & \simeq &
         1.1089 \times ({\rm cts/s}/\sq\arcsec) & \quad {\rm [STIS]} \ . \\
  \end{array}
\label{eq:intensity}
\end{equation}
Converting from luminosity density in
$L_\odot\ {\rm pc}^{-2}$ to surface brightness is done according to
$\mu_{814}/{\rm mag\ arcsec^{-2}}=
     25.712-2.5\,\log(I_{814}/L_\odot\,{\rm pc}^{-2})$
for F814W;
$\mu_{606}/{\rm mag\ arcsec^{-2}}=
     26.212-2.5\,\log(I_{606}/L_\odot\,{\rm pc}^{-2})$
for F606W; 
$\mu_{V,555}/{\rm mag\ arcsec^{-2}}=
     26.402-2.5\,\log(I_{V,555}/L_\odot\,{\rm pc}^{-2})$
for standard $V$ and F555W.

The ELLIPSE profiles extended to $R=12\farcs8\simeq48$~pc
from all images of the sample clusters, and thus in the model
fitting a constant-background term is included to allow for any errors in
sky subtraction in the automatic pipeline. Each profile was inspected, and
one or two very discrepant points in each of a handful of clusters were
excluded by hand.
Inspection of the data quality images for the WFC images showed that, due to
an unfortunate choice of CCD gain, the cores of several bright clusters
(B042, B063, B082, B147, and B151) were saturated, with raw intensities
above a nominal limit of 70 cts~s$^{-1}$~pix$^{-1}$ (28,000 cts~s$^{-1}$~arcsec$^{-2}$,
or $I_{F606W} \simeq\!23,600\ L_\odot\ {\rm pc}^{-2}$)
all the way out to $R\sim2\arcsec=7.5$~pc (or $\sim\!40$ pix) in most cases.
This is well beyond the
FWHM of the WFC PSF in either F606W or F814W; see below. Intensities as high
as this were not included in the model fitting, and thus the core
parameters of these five clusters are constrained only very indirectly.
None of the HRC or STIS images is saturated.

Table \ref{tab:M31sbprofs} gives the final, calibrated intensity profiles for all clusters. 
These are not corrected for extinction, which is discussed below. 
The reported F606W-
and F814W-band intensities are calibrated on the VEGAMAG scale, while the
STIS data are on the standard $V$ system.
The final column gives a flag for every point, which can
take one of four values. 
``BAD'' indicates that the intensity value is deemed dubious because
it strongly deviates from its neighbors or is obviously affected by nearby 
bright stars or image artifacts.
``SAT'' indicates that the isophotal intensity is above
the imposed saturation limit of 70 cts~s$^{-1}$~pix$^{-1}$
``DEP'' indicates that the radius is inside $R<2\ {\rm pix}=0\farcs1$ and the isophotal
intensity is dependent on its neighbors. The ELLIPSE output
includes brightnesses for 15 radii inside 2 pixels, but these are all measured
from the same 13 central pixels and are clearly not statistically independent.
To avoid excessive weighting of the central regions of the
cluster in the fits, only intensities at radii
$R_{\rm min}, R_{\rm min}+(0.5, 1.0, 2.0 {\rm px}), R>2.5\ {\rm px}$, were used.
``OK'' indicates that none of the above apply and the point is used in model fitting.

\subsection{Point-Spread Functions}
\label{subsec:PSF}

M31 clusters are clearly resolved with HST, but their observed core
structures are still affected by the PSF. We chose not to deconvolve
the data, instead convolving the structural models with a simple analytic 
description of the PSF before fitting.

To estimate the PSF for the WFC, ELLIPSE was used (again with circular symmetry
enforced) to produce intensity profiles out to radii of about $2\arcsec$
(40 pixels) for a number of isolated stars on a number of images, and combined
them to produce a single, average PSF. This was done separately for
the F606W and F814W filters.
We originally tried to fit these with simple Moffat profiles (with backgrounds
added), but found that a better description was given by a function of the form
\begin{equation}
I_{\rm PSF}/I_0 = \left[1+\left(R/r_0\right)^3\right]^{-\beta/3}\ .
\label{eq:psfform}
\end{equation}
The derived $r_0$ and $\beta$ values, and the implied FWHMs, are listed in Table
\ref{tab:psf}. 
Although the WFC PSF is known to vary over the instrument's field of view,
\citet{dean06} show that the model-fitting results are insensitive to this effect.

For HRC data, the same functional form (eq.\ref{eq:psfform}) was fitted to the
PSF images in F555W, F606W and F814W given by \citet{ak05} with
the results also listed in Table \ref{tab:psf}.
In the case of STIS, a similar procedure was followed \citep[see][]{hhhm02}
and it was found that a standard Moffat function,
$I_{\rm PSF}\propto [1+(R/r_0)^2]^{-\beta/2}$,
gives an adequate fit to
the PSF, with parameters also listed in Table
\ref{tab:psf}. These are the functions convolved with the structural
models before fitting to the observed intensity profiles.

\subsection{Extinction and Color Corrections}
\label{subsec:transmet}

The effective wavelengths of the ACS F606W and F814W filters are
$\lambda_{\rm eff}=5918$~\AA\ and $\lambda_{\rm eff}=8060$~\AA\ \citep{acs_cal}, 
so that from \citet{ccm89} $A_{606}\simeq 2.8\times E(B-V)$ and
$A_{814}\simeq 1.8\times E(B-V)$---both of which are also found by
\citet{acs_cal}. This implies, rather conveniently, that
$E(F606W-F814W) \simeq E(B-V)$. Individual reddening estimates---the sum of
foreground and M31-internal reddening---are available for most
of the clusters in the sample, derived using the procedure described by
\citet{b00}. The \citet{bh84} foreground value of
$E(B-V)=0.08$~mag is adopted for clusters without individual measurements.

\citet{acs_cal} give transformations from ACS F606W and F814W magnitudes to
standard $V$ and $I$ for both WFC and HRC measurements, including linear and
quadratic dependences on de-reddened $(V-I)_0$ color (their Tables 22 and
23). However, measured $(V-I)$ is not available 
for 5 of the 15 clusters in the present ACS sample, and thus
we chose instead to measure aperture $(F606W-F814W)$ colors from the current
data, and manipulate the \citeauthor{acs_cal} relations to find $(V-F606W)$ and
$(V-F814W)$ indices.

Comparing the various ``observed'' and ``synthetic'' transformations from
intrinsic $F606W$ magnitudes and $(V-I)_0$ colors to $V$ mags in
\citet{acs_cal}, it quickly becomes clear that there are systematic differences
between them, at the level of $\sim\!0.07$~mag in the color range
$0.8\la (V-I)_0\la 1.2$ appropriate for old clusters with sub-solar
metallicities. However, a simple linear relation,
\begin{equation}
(V-F606W)_0=-0.038 + 0.236 (V-I)_0
\label{eq:VIv606}
\end{equation}
both gives a good estimate of the average
of the \citeauthor{acs_cal} relations (for VEGAMAG F606W magnitudes) {\it and}
agrees very well with the predictions for old clusters in the
population-synthesis models of \citet{mar98,mar05}. [VEGAMAG $(V-F606W)_0$
and $(V-F814W)_0$ indices were kindly computed for us, as functions of cluster
age and metallicity, by C.~Maraston.]

To go further, VEGAMAG $F814W$ magnitudes must be related to $I$. 
All of the WFC and HRC transformations in \citet{acs_cal} imply that
$(F814W-I)_0\approx 0$ to within $\pm0.01$~mag or so, while the
population-synthesis models of \citeauthor{mar05} have a roughly constant
$(F814W-I)\simeq0.04$ for $0.8\la (V-I)_0\la 1.2$. Substituting
$(F814W-0.04)$ for $I$ in equation (\ref{eq:VIv606}) yields
\begin{equation}
(V-F606W)_0 = -0.04 + 0.31 (F606W-F814W)_0
\label{eq:acsv606}
\end{equation}
and then trivially,
\begin{equation}
(V-F814W)_0 = -0.04 + 1.31 (F606W-F814W)_0 \ ,
\label{eq:acsv814}
\end{equation}
both with estimated rms scatter of about $\pm0.05$~mag. These
conversions were applied to both WFC and HRC data, after correcting the
calibrated $F606W$ and $F814W$ magnitudes/intensities for extinction. 

Calibrating the other observations onto the standard $V$ scale
was more straightforward. The ACS/HRC F555W filter is similar to
the standard $V$ filter, and the data for G001 were calibrated using
the extinction ratio and transformation given by
\citet{acs_cal} (their Tables~15 and 23).
The data for the 19 STIS clusters are already calibrated on the
standard $V$ scale \cite{hhhm02}, as are the WFPC2 data from \citet{bh01}.
The reddening values used to correct for
extinction are recorded in Table \ref{tab:m31meas}.

Given the color transformations above, the measured ACS aperture colors 
plus the assumed reddening values allow us to predict a reddened $V-I$ color
to be compared with ground-based aperture colors.
For the 10 of 15 ACS-observed clusters with $(V-I)$ colors
\citep[ground-based, mainly from the compilation in][]{b00},
the agreement between predicted and observed $(V-I)$ 
colors is very good: the mean offset is $0.02\pm0.02$~mag, with a
median absolute deviation of 0.06~mag.
The cluster B023 has been observed in all three bandpasses discussed here: F606W
and F814W with ACS/HRC, and $V$ with STIS. The fits of \citet{kin66} models
to these three independent profiles in \S\ref{sec:fit_results} yield
extinction-corrected total magnitudes of $F814W=12.09$, $F606W=12.88$, and
$V=12.99$, with uncertainties of about $\pm0.02$ mag in each. The implied
global $(F606W-F814W)_0=0.79$ is essentially identical to the
aperture color computed from the measurements in Table~\ref{tab:m31meas}, and the
fitted global $(V-F606W)_0$ and $(V-F814W)_0$ are within 0.1 mag of the
predicted values---suggesting a level of agreement similar
to that between $(V-I)_{\rm pred}$ and $(V-I)_{\rm obs}$ more generally.

The cluster B041 is somewhat unusual. Its integrated colors
suggest that it is perfectly average, compared to the rest of the M31 cluster
system. However, the ACS image shows that a very bright, red star about
0\farcs5 from the cluster center is responsible for the redness of its
$(F606W-F814W)$ and $(V-I)$ aperture colors.
Both the innermost intensities from the ELLIPSE profiles and the total cluster $F606W$ and
$F814W$ magnitudes from model fitting (in which the star is ``smoothed
over'' in some sense), imply a de-reddened
$(F606W-F814W)_0\simeq0.4$ ---much bluer than any other cluster in the sample.
In fact, the population-synthesis model predictions of
C.~Maraston suggest that the oldest possible age for such a blue cluster is
$\lesssim 1.5$~Gyr. Because of its red integrated color, this cluster is not
included in any of the `massive young M31 cluster' samples of \citet{fp05} ---
showing that integrated properties are not always
reliable guides to the true nature of extragalactic globular clusters.
There are eight clusters in the WFPC2 sample which are members of
`blue luminous compact cluster' samples `A' or `B'  in \citet{fp05}.
These are suspected of being young clusters by virtue of very blue colors
[$(B-V)_0<0.45$] or strong H$\beta$ absorption. These clusters and B041
are not included in any of the fundamental plane correlations in \S\ref{subsec:fpcor};
a total of 84 M31 clusters are considered there.

\section{Models}
\label{sec:models}

\begin{figure}
\plotone{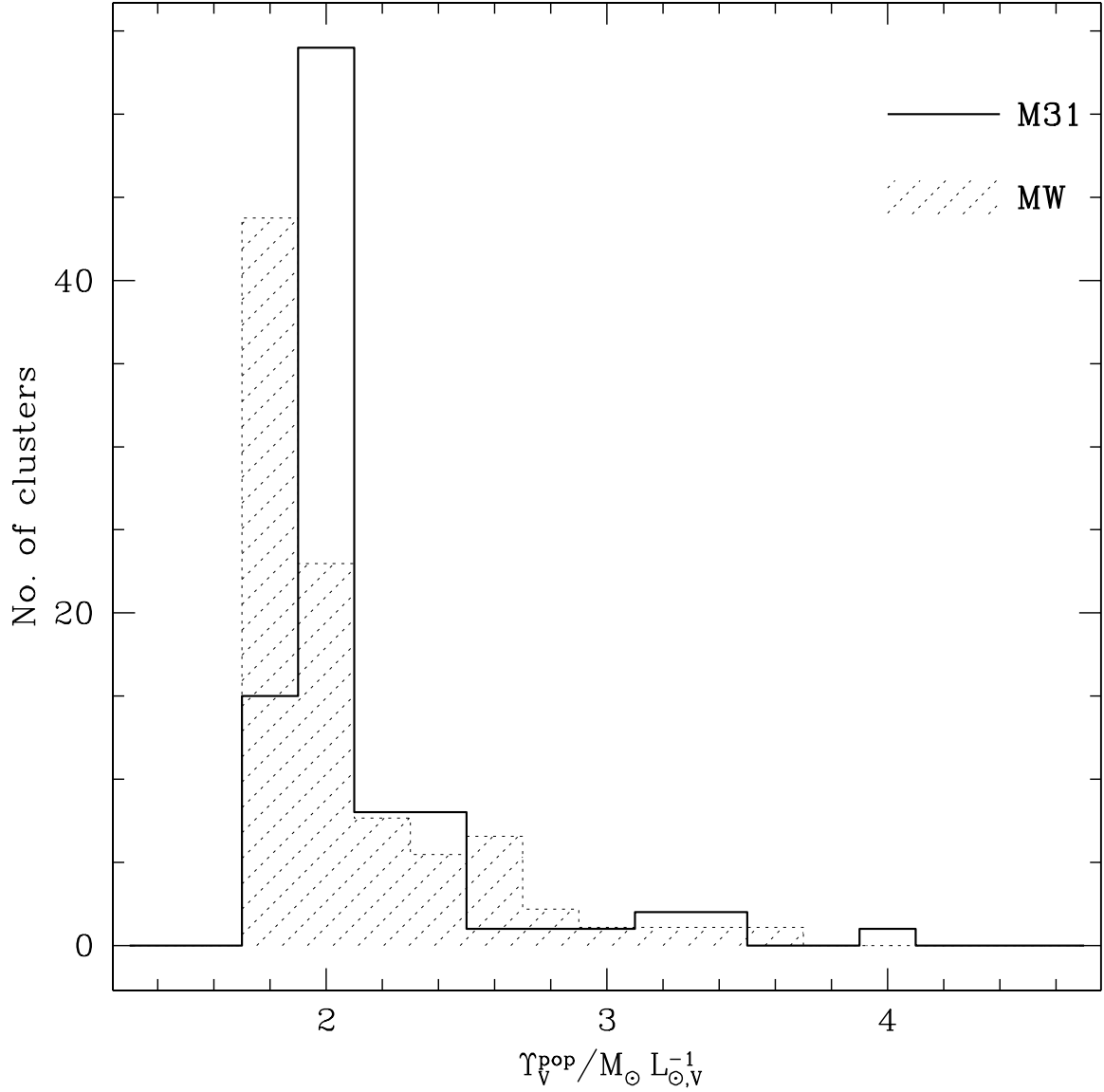}
\caption{
Histogram of population-synthesis $V$-band mass-to-light ratios for 
M31 GCs assuming an age of 13 Gyr; and the same for Galactic globulars, from
\citet{mv05}.
\label{fig:m31mtol}
}
\end{figure}

Converting luminosity-based parameters from the model fits 
into mass-based quantities requires knowledge of mass-to-light ratios.
Velocity-dispersion measurements do exist for a fraction of the sample (see \S\ref{subsec:apvel})
but the measurements are far from complete or uniform.
Instead, as in \citet{dean06}, we use population-synthesis
models to {\it predict} mass-to-light ratios for the full sample,
using the code of \citet{bc03} and assuming the (disk) stellar IMF of \citet{chab03}.
These can then be used to produce (for example) predicted velocity dispersions that can be
compared to current and future spectroscopic data.
The values of $\Upsilon_V^{\rm pop}$ for an assumed age of 13 Gyr
are discussed further and tabulated in \S\ref{subsec:deriv_quant}
and shown as a histogram in Figure \ref{fig:m31mtol}.
For comparison, also shown is the (re-normalized to the same sample size) 
distribution of mass-to-light ratios for 85 
Galactic globulars with parameters cataloged by \citet{mv05}, calculated using the same population-synthesis
model code with the same assumed stellar IMF and for the same assumed age of
13 Gyr as in the present analysis. The distribution peaks at a
slightly higher value of $\Upsilon_V^{\rm pop}$ in M31 vs.~the Milky Way,
which simply reflects a slightly higher average GC metallicity in the sample
of M31 clusters.

The cluster structural models used in the fits are described in detail 
by \citet{dean06}. The three models considered here are the often-used
\citet{kin66} single-mass, isotropic, modified isothermal sphere; 
an alternate modification of a single-mass, isotropic isothermal
sphere, based on a model originally introduced by \citet{wil75} for elliptical
galaxies; and the \citet{sersic} or $R^{1/n}$ model.
\citet{wil75} models are similar to the standard \citet{kin66} models,
in that their cores are nearly isothermal, but their envelope structures are
relatively more extended---a feature that has been found to better describe the
density profiles of old globulars and young massive clusters in the  Local
Group and in NGC~5128. See \citet{mv05} and \citet{dean06} for more detailed
discussion and for general comparisons between the two types of models.
Two additional models, `power-law' profiles and the analytic \citet{king62} model, are 
found by \citet{dean06} to provide little or no information beyond that contained in
the first three models, and are not discussed here.

Fitting the structural models to the data involves first 
selecting a value for $r_0$, the scale radius,\footnote{%
The definition of $r_0$ varies between models: for \citet{sersic} models,
$I(r_0) \equiv 0.5 I_0$, while for \citet{kin66} and \citet{wil75}
this relation is only approximate. See \citet{dean06} for details.}
and computing the 
dimensionless model profile $\widetilde{I}_{\rm mod}\equiv I_{\rm mod}/I_0$.
The model is then convolved with the appropriate PSF 
having parameters taken from Table \ref{tab:psf} to yield
the product $\widetilde{I}_{\rm mod}^{*} (R | r_0)$. The fitting procedure allows
for a varying scale radius $r_0$ and non-zero background $I_{\rm bkg}$, and so minimizes
\begin{equation}
\chi^2  =
  \sum_i{
 \frac{\left[I_{\rm obs}(R_i)
             - I_0\times \widetilde{I}_{\rm mod}^{*}(R_i | r_0)
             - I_{\rm bkg}\right]^2}
      {\sigma_i^2}
        }
\label{eq:chi2}
\end{equation}
for the measured intensity profile $I_{\rm obs}(R_i)$ and uncertainties $\sigma_i$ of each cluster in
Table \ref{tab:M31sbprofs}.

Uncertainties on the fitted and derived model parameters are estimated in 
the usual way, from the range of their values in all
models for which $\chi^2$ is within some specified distance of the absolute
minimum (e.g., $\Delta\chi^2\le 1$ for 68\% confidence intervals).
Because the formal error bars estimated by ELLIPSE for the isophotal intensities are
artificially small, the best-fit $\chi_{\rm min}^2$ can be exceedingly
high ($\gg\!N_{\rm pts}$) even when a model fit is actually very good. 
This would result in unrealistically small estimates of parameter uncertainties. 
We chose to re-scale the $\chi^2$ for all fitted models by a common factor chosen to make
the global minimum $\chi_{\rm min}^2 = (N_{\rm pts}-4)$, where $N_{\rm pts}$
is the number of points used in the model fitting. Under this re-scaling, the global 
minimum reduced $\chi^2$ per degree of freedom is exactly one.
This prescription for deciding fit uncertainties does not affect the identification 
of the best-fitting model itself, since the {\it relative} sizes of the uncertainties on
individual data points are not changed.

\section{Results}
\label{sec:fit_results}

\subsection{Model-fitting}

\begin{figure}
\plotone{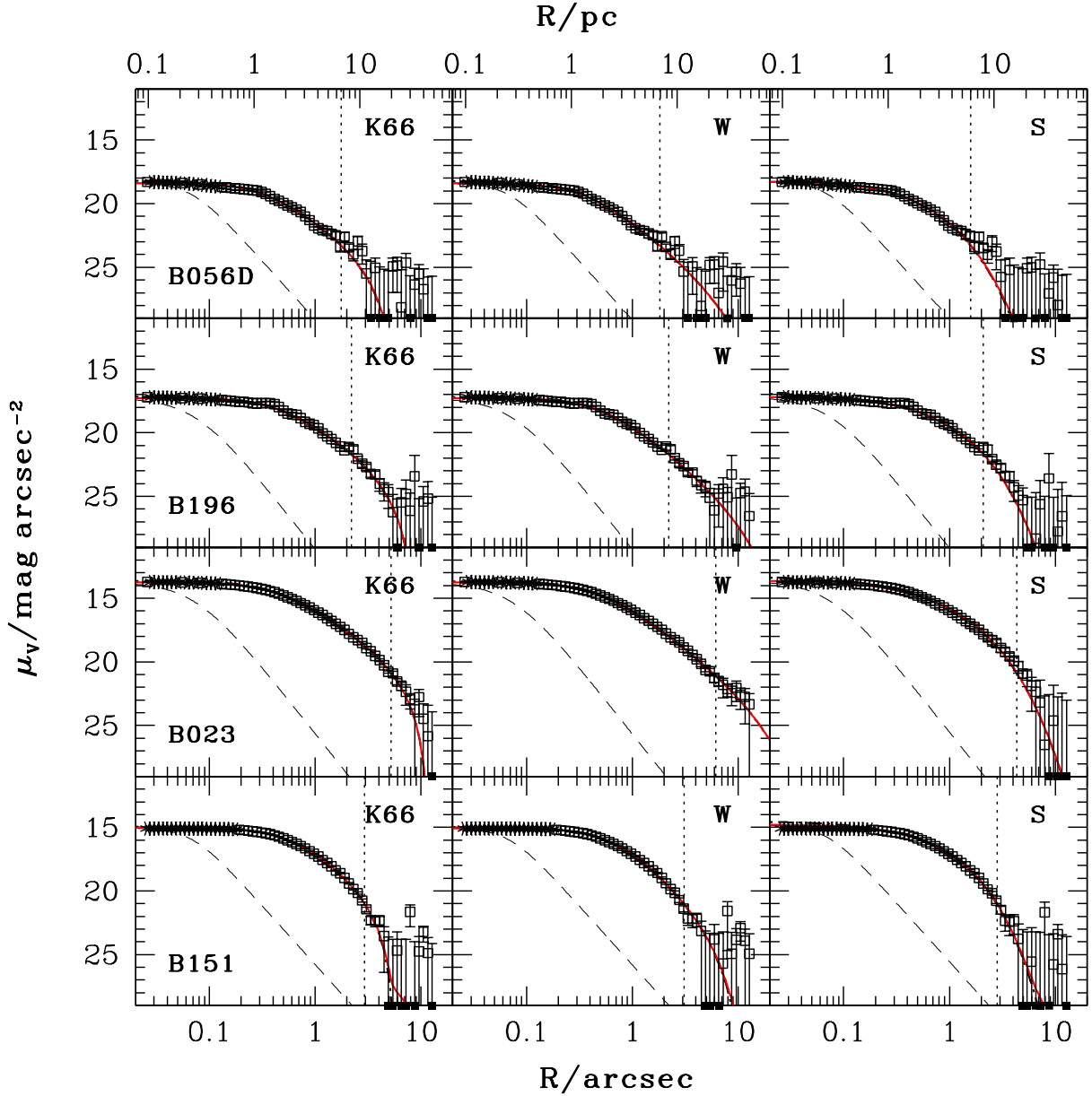}
\caption{
Surface brightness profiles and model fits to four M31 GCs. Clusters
are in order of increasing luminosity, from top to bottom: B056D, B196,
B023, and B151. The three panels for each cluster are, from left to right: \citet{kin66},
\citet{wil75} and \citet{sersic} models.
Dashed curves trace the PSF intensity profiles and solid (red)
curves the PSF-convolved best-fit models with added background.
Vertical dotted lines mark the 
radius where the best-fit cluster intensity is equal to the background.
Open squares are ELLIPSE data points included in the least-squares model fitting
and the asterisks are points (flags of BAD, SAT, or DEP in Table \ref{tab:M31sbprofs}) 
not used to constrain the fits. 
\label{fig:exfitM31}
}
\end{figure}

Figure \ref{fig:exfitM31} shows example surface brightness profiles and model fits 
for four M31 clusters chosen to span the observed luminosity range.
The observed data plotted have had the {\em fitted} $I_{\rm bkg}$ subtracted
and then been converted to (extinction-corrected) surface brightnesses.
Points with intensities below the subtracted background in the upper panel are
represented here as solid points placed on the lower $x$-axis, with error bars
extending upwards. The Figure shows that the M31 GCs are well-resolved by
the HST imaging, and also demonstrates the core saturation for the brightest
clusters exemplified by B151. For these particular clusters there is relatively little
difference in the goodness-of-fit of the three different models. As expected,
the \citet{wil75} models, with their extended halos, favor somewhat lower background
levels, and the `cuspier' \citet{sersic} models favor higher central surface brightnesses.
The clusters have projected half-light radii $R_h=2-4$~pc and King-model
concentrations in the range 1--1.5, quite typical values for GCs in both M31 \citep{bhh02}
and the Milky Way \citep{h96}. 

Table \ref{tab:m31fits} summarizes the basic ingredients of all model fits to the
full sample. The first column
in this table gives the cluster name; the second, the detector/filter
combination from which the density profile was derived; 
the third, the extinction in the F606W band;
the fourth, the color term to transform photometry from the native bandpass of
the data to the standard $V$ scale (see \S\ref{subsec:transmet}),
and the fifth column lists the number of points in the intensity profile
that are flagged as OK in Table \ref{tab:M31sbprofs} above, and thus were
used to constrain our model fits.
Subsequent columns in Table \ref{tab:m31fits} cover three lines for each
cluster, one line for each type of model fit. Column (6) identifies the model.
Column (7) gives the minimum unreduced $\chi^2$ obtained for that class
of model (without the re-scaling applied for uncertainty estimation). 
Column (8) gives the best-fit background intensity.
Column (9) gives the dimensionless central potential $W_0$ of the best-fitting
model (for \citealt{kin66} and \citeauthor{wil75} models only).
Column (10) gives the concentration $c\equiv\log(r_t/r_0)$, or
the \citeauthor{sersic} index $n$ of the best fit.
Column (11) gives the calibrated and extinction-corrected central surface
brightness in the native bandpass of the data.
Column (12) gives the logarithm of the best-fit scale radius $r_0$ in 
arcsec (see \S\ref{sec:models}), and Column (13) gives the logarithm of $r_0$
in units of pc (obtained from the angular scale assuming $D=780$~kpc for M31). 
Error bars on all these parameters were defined in the same way as \citet{dean06}.

\subsection{Derived quantities}
\label{subsec:deriv_quant}

Table \ref{tab:m31phot} contains a number of other structural cluster
properties derived from the basic fit parameters. The details of their
calculation are given in \citet{dean06}; the contents are:
\begin{enumerate}
\item $\log\,r_t=c+\log\,r_0$, the model tidal radius in pc (always
infinite for \citeauthor{sersic} models)
\item $\log\,R_c$, the projected core radius of the model
fitting a cluster, in units of pc
\item $\log R_h$, the projected half-light, or effective, radius of a model:
that radius containing half the total luminosity in projection, in units of pc%
\footnote{Calculations show that the three-dimensional half-light radius $r_h=4/3 R_h$.}
\item $\log\,(R_h/R_c)$, a measure of cluster concentration 
\item $\log\,I_0$, the logarithm of the best-fit central ($R=0$) luminosity surface
density in the $V$ band, in units of $L_{\odot,V}\,{\rm pc}^{-2}$
\item $\log\,j_0$, the logarithmic central ($r=0$) luminosity volume density in the $V$ band
(for \citeauthor{sersic} models the density at the 3-D radius $r_0$), 
in units of $L_\odot\,{\rm pc}^{-3}$
\item $\log\,L_V$, the logarithm of the total integrated model
luminosity in the $V$ band
\item $V_{\rm tot}=4.83-2.5\,\log\,(L_V/L_\odot)+ 5\,\log\,(D/10\,{\rm pc})$
is the total, {\it extinction-corrected} apparent $V$-band magnitude of a
model cluster
\item $\log I_h \equiv \log\,(L_V/2\pi R_h^2)$ is the $V$-band
luminosity surface density averaged over the half-light/effective radius, in
units of $L_{\odot,V}\,{\rm pc}^{-2}$.
\end{enumerate}
The uncertainties on all of these derived parameters have been
estimated (separately for each given model family) following the $\chi^2$
re-scaling procedure described above.
The re-computed values for the 59 clusters from \citet{bhh02}, for \citet{kin66}
models only, are reported at the end of this Table.

Table \ref{tab:m31mass} next lists a number of ``dynamical'' cluster
properties derived from
the structural parameters already given  plus a mass-to-light ratio.
The first two columns of this table contain the name of each cluster and the
combination of detector/filter for the observations. Column
(3) lists the mass-to-light ratio, in solar units, adopted for
each object from the analysis in \S\ref{sec:models}.
The values of $\Upsilon_V^{\rm pop}$ assume an age of 13 Gyr for all clusters. 
The error bars on $\Upsilon_V^{\rm pop}$ in Table \ref{tab:m31mass} include
a $\pm 2$-Gyr uncertainty in age, as well as the previously tabulated
uncertainties in [Fe/H]. The remaining entries in Table \ref{tab:m31mass}
are, for the best fit of every model type to every cluster:
\begin{enumerate}
\item $\log\,M_{\rm tot}=\log\,\Upsilon_V^{\rm pop}+\log\,L_V$, the
integrated model mass in solar units, with $\log\,L_V$ taken from Column (10)
of Table \ref{tab:m31phot}
\item $\log\,E_b$, the integrated binding energy in ergs, defined through
$E_b\equiv -(1/2)\int_{0}^{r_t} 4\pi r^2 \rho \phi\,dr$
\item $\log\,\Sigma_0$, the central surface mass density in units of $M_\odot\,{\rm pc}^{-2}$
\item $\log\,{\rho}_0$, the central volume density in units of $M_\odot\,{\rm pc}^{-3}$
(for \citeauthor{sersic} models the density at the 3-D radius $r_0$)
\item $\log\, \Sigma_h$, the surface mass density averaged over the effective radius $R_h$,
in units of $M_\odot\,{\rm pc}^{-2}$
\item $\log\,\sigma_{{\rm p},0}$, the predicted line-of-sight velocity
dispersion at the cluster center (or at $R_c$ for \citeauthor{sersic} models), in km~s$^{-1}$
\item $\log\,v_{{\rm esc},0}$, the predicted central ``escape'' velocity in
km~s$^{-1}$
\item $\log\,t_{\rm rh}$, the two-body relaxation time at the model
projected half-mass radius, in years
\item $\log f_0\equiv \log\,\left[\rho_0/(2\pi \sigma_c^2)^{3/2}\right]$,
a measure of the model's central (at $r_0$ for \citeauthor{sersic} models) 
phase-space density or relaxation time, in units of
$M_\odot\,{\rm pc}^{-3}\,({\rm km\ s^{-1}})^{-3}$.
For $f_0$ in these units, and $t_{rc}$ in years, $\log t_{rc} \simeq 8.28 - \log f_0$ 
\citep{mv05}.
\end{enumerate}
The uncertainties in these derived dynamical quantities are estimated from
their variations around the minimum of $\chi^2$ on the model grids,
as above, combined in quadrature with the uncertainties in the
population-synthesis model $\Upsilon_V^{\rm pop}$.

Finally, Table \ref{tab:m31kappa} provides the last few parameters
required to construct the fundamental plane of globular clusters in M31
under any of the equivalent formulations of it in the literature (see
\S\ref{subsec:fpcor} below). The last three columns of Table \ref{tab:m31kappa} 
are modified versions of the ``$\kappa$'' parameters of \citet{bbf92},
who found that early-type galaxies and (separately) globular clusters
defined the edges of planes in $\kappa$ space. We define the parameters
using mass rather than luminosity surface density, as this is more useful
for comparing globulars to younger clusters and galaxies \citep[see][]{dean07},
and emphasize this by using the notation $\kappa_{m,i}$.
\begin{equation}
\begin{array}{rcl}
\kappa_{m,1} & \equiv & (\log\,\sigma_{{\rm p},0}^2 + \log\,R_h)/\sqrt{2} \\
\kappa_{m,2} & \equiv & (\log\,\sigma_{{\rm p},0}^2 + 2\,\log\Sigma_h
                      - \log\,R_h)/\sqrt{6}    \\
\kappa_{m,3} & \equiv & (\log\,\sigma_{{\rm p},0}^2 - \log\,\Sigma_h
                      - \log\,R_h)/\sqrt{3} \\
\end{array}
\label{eq:kspace}
\end{equation}
In calculating $\kappa_{m,1}$, $\kappa_{m,2}$, and $\kappa_{m,3}$ for Table
\ref{tab:m31kappa}, the $\sigma_{{\rm p},0}$ predicted
in Table \ref{tab:m31mass} (Column 7) by adoption of
population-synthesis mass-to-light ratios has been used, and $\Sigma_h$ evaluated
by adding $\log\,\Upsilon_V^{\rm pop}$ to $\log\,I_h$ from Column
(12) of Table \ref{tab:m31phot}. (This means $\kappa_{m,3}$ is therefore
{\it independent} of the assumed mass-to-light ratio.) $R_h$ is also taken
from Table \ref{tab:m31phot} but put in units of kpc rather than pc, for
compatibility with the galaxy-oriented definitions of \citet{bbf92}.

\subsection{Predicted Velocity Dispersions}
\label{subsec:apvel}

One use for the material presented in the previous subsections is to predict
observable, line-of-sight velocity dispersions averaged over circular
apertures of any radius, which can then be compared directly to extant and
forthcoming spectroscopic data on the clusters in the
sample. Ultimately such comparisons will be useful not only for independent
assessments of the self-consistency of the structural model fits, but
also to evaluate how close the dynamical mass-to-light ratios in these old
globulars come to the expectations of population-synthesis models with no dark
matter.

Given any of the models with a fitted $r_0$ and $c$ or $n$ for any 
cluster, solving Poisson's and Jeans' equations and projecting along 
the line of sight yields a dimensionless velocity-dispersion 
profile, $\widetilde{\sigma}_{\rm p}=\sigma_{\rm p}(R)/\sigma_{{\rm 
p},0}$ as a function of projected clustercentric radius $R$. The 
density-weighted average of $\widetilde{\sigma}_{\rm p}^2$ within a 
circular region of radius $R_{\rm ap}$ then gives a dimensionless 
aperture dispersion for the cluster. Finally, normalizing with the 
predicted $\sigma_{{\rm p},0}$ from the fits in Table~\ref{tab:m31mass}, we can 
predict an observable $\sigma_{\rm ap}(R_{\rm ap})$ for every model 
for every cluster, and for any specified $R_{\rm ap}$. Table~\ref{tab:m31vels} gives 
the results of these calculations for each of 5 apertures: $R_{\rm 
ap}=R_h/8, R_h/4, R_h, 4\,R_h$, and $R_{\rm ap}=8\,R_h$. For a 
distance of 780 kpc to M31, a typical GC $R_h=3$--4 pc translates
to roughly 0\farcs8--1\farcs1, so the range of apertures over which 
$\sigma_{\rm ap}$ is calculated should be a good match to reasonable
observations. The velocity dispersion inside other apertures
can be obtained by interpolation.

\begin{figure}
\plotone{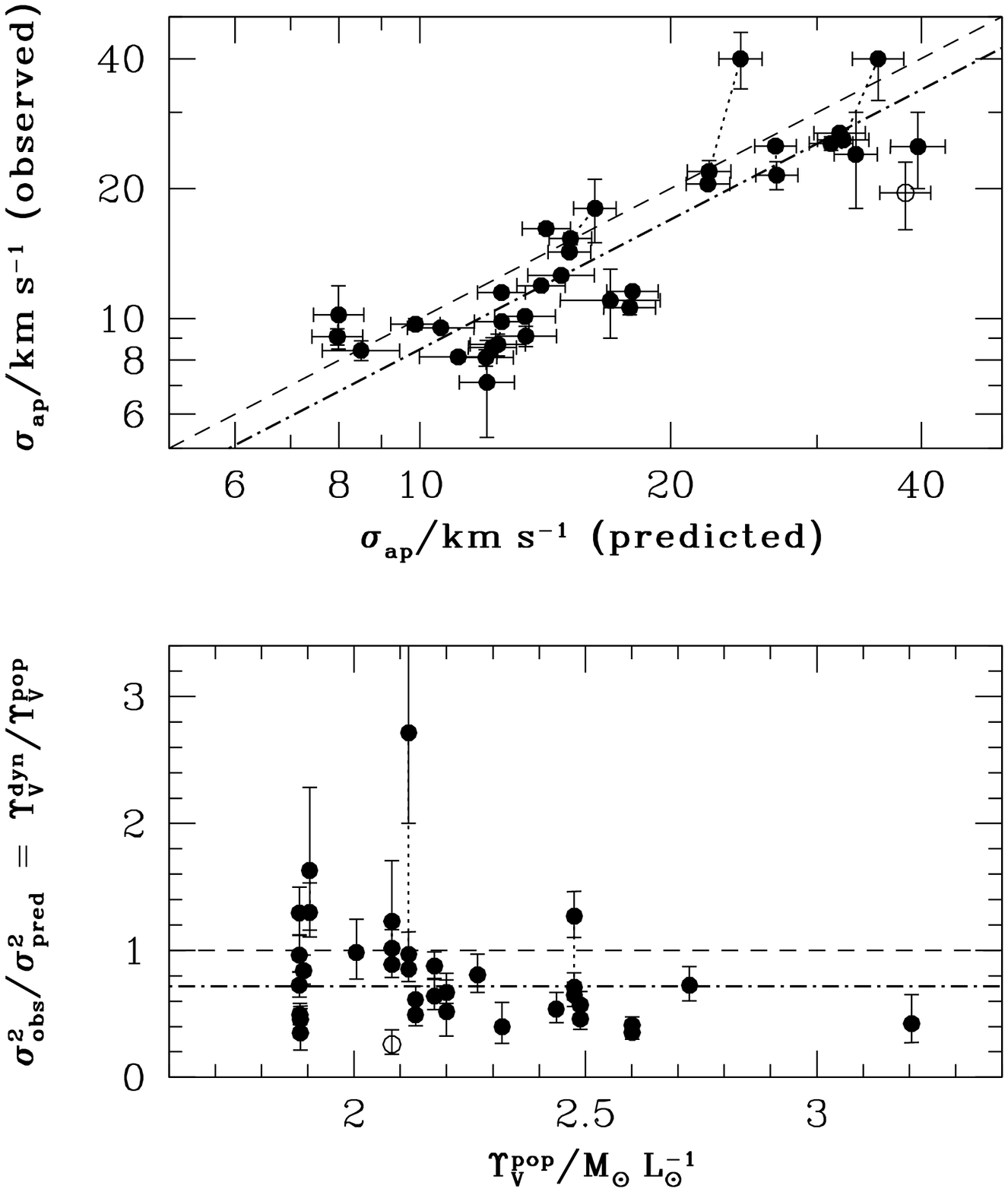}
\caption{
Top: comparison of structural-model predictions for central velocity
dispersion with observations from \citet{djo97}, \citet{dg97},
and \citet{pet89}. Open symbol is cluster B037, discussed in \S\ref{subsec:indiv}.
Points connected by short-dashed lines are duplicate observations of the same cluster.
Long-dashed line is line of equality; dot-dashed line is median ratio 
${\sigma}_{\rm obs}/{\sigma}_{\rm pred}=0.85$.
Bottom: comparison of dynamical and population-synthesis derived mass-to-light
ratios. Long-dashed line is line of equality; dot-dashed line is median ratio 
${\Upsilon}^{\rm dyn}_V/{\Upsilon}^{\rm pop}_V=0.725$.
\label{fig:veldisp}
}
\end{figure}

These predicted velocity dispersions can already be compared 
in a limited way to observations.
Measured velocity dispersions for 29 M31 GCs, of which 22 are included in
the current {\it HST} sample,
are available from from \citet{cohen06}, \citet{dg97}, \citet{djo97}, and \citet{pet89}.
These are summarized in Table~\ref{tab:m31veldisp}.
Figure~\ref{fig:veldisp} compares observed and model-predicted
velocity dispersions, where the model predictions are derived from
interpolation on the values in Table~\ref{tab:m31vels} to the aperture
size reported for the observations.
Aperture sizes were computed as $R_{\rm ap} = \sqrt{l w/\pi}$
where $l$ and $w$ are the reported spectrograph slit dimensions.

Figure~\ref{fig:veldisp} shows that, except for a constant ratio, there is 
generally good agreement between observed and predicted velocity dispersions 
(the exception, cluster B037, is discussed further in \S\ref{subsec:indiv}). 
This justifies our use of population synthesis model 
mass-to-light ratios to derive masses and other properties such as binding energy. 
This procedure gives us a much larger sample 
of clusters to work with for a wider variety of analyses than would be possible
using only direct velocity dispersion measurements. The median ratio
between observed and predicted velocity dispersions is 
${\sigma}_{\rm obs}/{\sigma}_{\rm pred}=0.85$, with an inter-quartile range
$\pm0.15$. 
This corresponds to an ratio between dynamical and population-synthesis-derived 
mass-to-light ratios of ${\Upsilon}^{\rm dyn}_V/{\Upsilon}^{\rm pop}_V=0.73\pm0.25$,
consistent with the value for this ratio of $0.82\pm 0.07$ found by \citet{mv05}
for Milky Way and old Magellanic Cloud clusters.
The lower dynamical mass-to-light ratios compared to those derived from
population-synthesis models is consistent with with theoretical expectations
that old Milky Way and M31 GCs have lost low-mass stars through evaporation,
an effect not included in the models.

\subsection{Comparison of Fits in F606W and F814W}
\label{subsec:bandcomp}

\begin{figure}
\plotone{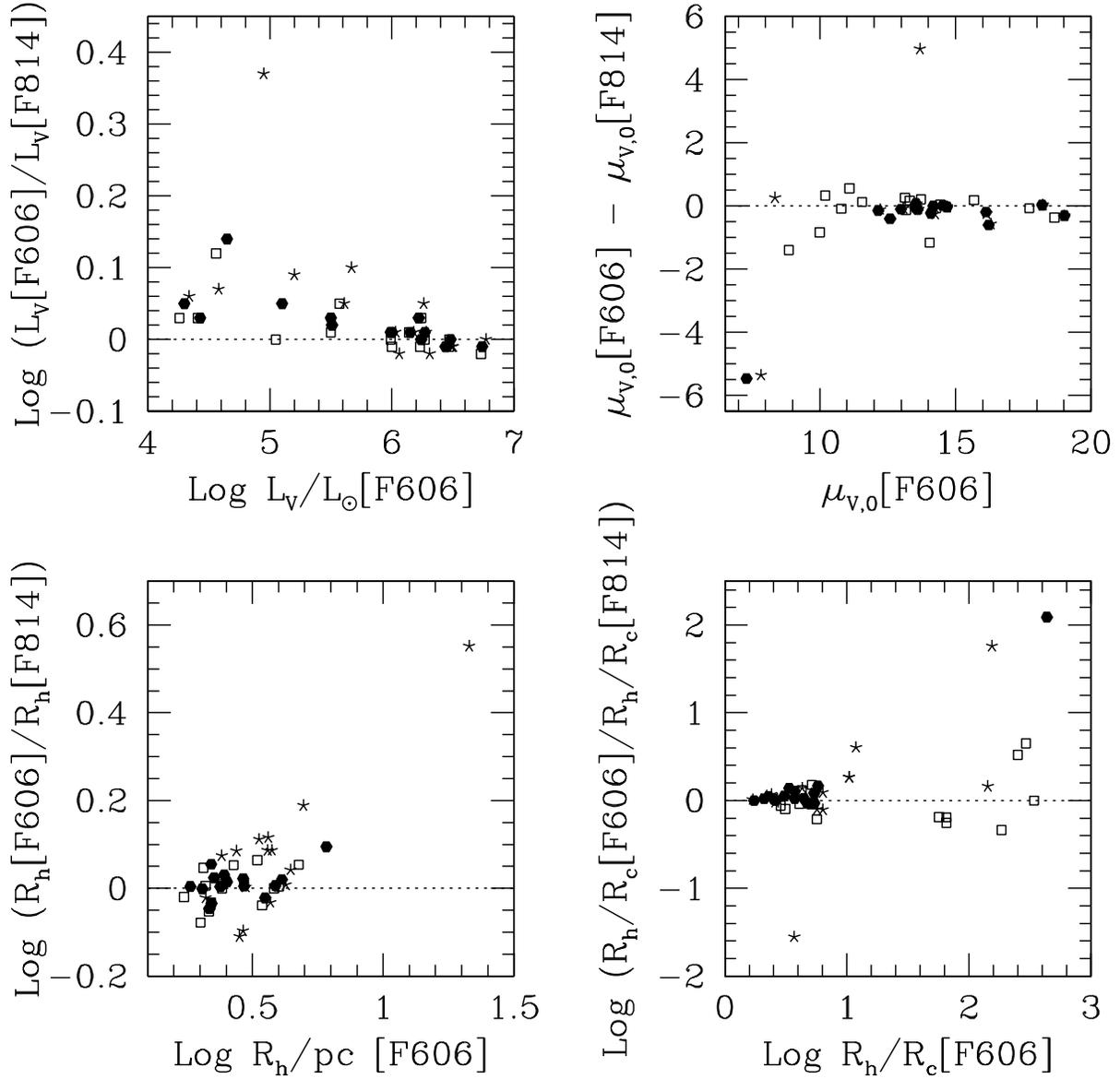}
\caption{
Comparison of parameters for  model fits to 15 M31 clusters
observed in both F606W and F814W filters. 
Top left: model total magnitude,
top right: central surface brightness,  
bottom left: projected half-light radius, and
bottom right: ratio of half-light to core radii (a measure of concentration).
Filled circles: \citet{kin66} models, 
squares: \citet{sersic} models,
stars: \citet{wil75} models.
\label{fig:filtcomp}
}
\end{figure}

Comparing model fits to the same cluster observed in different filters allows 
assessment of the systematic errors and color dependencies in the fits. 
Figure~\ref{fig:filtcomp} compares the
parameters derived from fits to the 15 ACS clusters observed in both F606W and F814W.
In general the agreement is quite good, with somewhat higher scatter for the
\citet{wil75} and \citet{sersic} model fits than for the \citet{kin66} fits.
There are a few unsuccessful fits with very high concentrations, small sizes
and extremely bright central surface brightnesses: the \citet{kin66} model fit to B082 in F606W; 
\citet{wil75} model fit to B042 in F606W and B063 in F814W; and the \citet{sersic} model
fit to B147 in F814W. These are all clusters in which the central core data are
not used due to saturation; fortunately  the data in the other filter produce
a more reasonable-looking fit. For all three models, the central $V$-band surface brightnesses
as fit to the F814W data are slightly fainter than those from the
F606W data, while the total model luminosities are essentially the same.
The latter implies that our estimated $(V-F606W)$ and $(V-F814W)$ colors in
\S\ref{subsec:transmet} are correct, while the former implies that the cluster
centers are slightly bluer than the average cluster color. This could indicate
the presence of blue stragglers or blue horizontal branch stars
in these massive clusters, but it could also
be due to the differing levels of saturation. 
A broader range of data will be needed
to investigate the issue of color gradients in M31 GCs more thoroughly
\citep[for a discussion of color gradients in Milky Way clusters, see][]{dp93}.

Because the model fit results in the two ACS bands are quite similar, and also
because for the STIS and WFPC2 data we have only $V$-band data, we do
not consider the F814W model fits in the following analyses. 
Parameters derived from the model fits given in Tables~\ref{tab:m31phot}--\ref{tab:m31vels} 
refer to the F606W band observations only. The exception
is B082, where we substitute the F814W fit results for the \citet{kin66} fit.

\subsection{Comparison to previous results for individual clusters}
\label{subsec:indiv}

The famous M31 cluster G001 has discrepant published values of its half-light
radius \citep[see][]{bhh02,ma06a}, which was a major motivation for re-analyzing
it in this paper. 
\citet{mey01} reported a three-dimensional $r_h=3\farcs7$, which converts to a projected 
$R_h=2\farcs8$, while \citet{bhh02} measured $R_h=0\farcs82$ and \citet{cmd4}
$R_h=0\farcs70$.%
\footnote{
\citet{lar01c} pointed out that the integrated photometry given by \citet{mey01}
implies a half-light radius of about 1\farcs18, much closer to
other values.}
Our \citet{kin66} model fits to the ACS/HRC F555W data resulted in a
half-light radius $R_h=0\farcs85$, and our \citet{wil75} fit to the same data
yielded $R_h=1\farcs2$. The \citet{mey01} value is clearly the outlier,
likely because it is for the half-{\it mass}, rather than the half-light radius,
and because the \citet{mey01} analysis used multi-mass \citet{kin66} models.
The surface brightness profile from the HRC observations is almost identical
to that presented in \citet{mey01}, which combined data from WFPC2 observations
including those used by \citet{cmd4}, so we conclude that the physical model
assumptions used are responsible for the differences between authors. 
Recently, \citet{ma07} also analyzed the ACS/HRC data used here: these authors derived
a higher concentration $c=2.01\pm0.02$ compared to our \citet{kin66} model value of 1.77,
and a value of $R_h=1\farcs73\pm0\farcs07$ about twice as large as ours. 
This appears to be the result of weighting
the outermost points in the surface brightness profile more heavily, 
yielding a larger tidal radius and higher concentration.
The dashed line in \citeauthor{ma07}'s Figure~3, which represents their 
model fit with the outermost surface brightness points excluded, is 
quite similar to our model fit, and we believe this to be more robust
particularly in the better-defined inner part of the cluster.

The heavily-reddened M31 cluster B037 has also been the subject of
intensive study. \citet{ma06a} analyzed the same F606W image of B037 discussed
here, fitting a \citet{king62} model to a surface brightness profile made
from a PSF-deconvolved image. They derived a core radius $r_0=0\farcs72$, half-light
radius $r_h=1\farcs11$, concentration $c=0.91$,
and central surface brightness ${\mu}(0) =17.21$~mag~arcsec$^{-2}$ in F606W
[using our values for extinction and $V-F606W$ color, this becomes $\mu_{V,0}=13.49$.]
Our model fits generally result in a somewhat higher concentration
and smaller core radius [$r_0=0\farcs56$, $r_h=1\farcs09$, $c=1.23$ for the \citet{kin66} model].
This relatively small change would not likely have affected \citeauthor{ma06a}'s
conclusions about the nature of this cluster.

\citet{ma06b} predict a velocity dispersion of 72~km~s$^{-1}$ 
for B037, on the basis of a mass estimate of $\approx 3 \times10^7 M_{\odot}$. 
However, these $M$ and $\sigma$ 
values are based on a population-synthesis $M/L_V$ ratio applied to an 
intrinsic cluster luminosity obtained by assuming a distance of $\sim950$~kpc 
to M31, instead of our adopted 780~kpc (and a large $E(B-V)$ close to
the one we have used). Thus, comparison of their numbers with 
ours requires first multiplying their derived mass by $(780/950)^2 = 2/3$. 
In addition, the \citeauthor{ma06b} mass and velocity dispersion are actually 
calculated for what the cluster would have been at an age of 10~Myr. 
However, the same population-synthesis models that they (like us) 
used---i.e., \citet{bc03}---show that the cumulative effects 
of supernova explosions, massive-star winds, and AGB mass loss over 12--13~Gyr 
lead to a reduction of the 10-Myr total cluster mass by another factor 
of $\simeq 2$ at its current, old age \citep[see also, e.g.,][]{frv97}. 
Therefore, the {\it present-day} mass of B037 
implied by the work of \citeauthor{ma06b} is in fact $\approx 1 \times 10^7 M_{\odot}$,
in good agreement with our results here (Table~\ref{tab:m31mass}).

The high velocity dispersion predicted for B037 by \citeauthor{ma06b} is a `global' 
(not central) value following from the virial theorem assuming both their 
high mass and a rough estimate of the cluster half-light radius ($\simeq 
2.6$~pc in their analysis), which is significantly smaller than that in our 
more accurate profile fitting ($R_h=3.9$~pc; e.g., Table~\ref{tab:m31vels}, column 5). 
Scaling to a 3-times smaller present-day mass, as above, and to a 
$\sim1.5$-times larger $R_h$, the expected global velocity dispersion following 
self-consistently from Ma et al. is 72~km~s$^{-1} \times \sqrt{(1/3)/1.5} = 34$~km~s$^{-1}$, 
which is then entirely consistent with our predicted $\sigma_{\rm ap}$ 
inside the largest apertures in Table~\ref{tab:m31vels}.

\citet{cohen06} has {\it measured} a velocity dispersion of $19.6\pm3.5$~km~s$^{-1}$
for B037 within an aperture of $R_{\rm ap}\simeq1.6$~arcsec. Our predicted 
dispersion within this aperture follows from interpolation on Table~\ref{tab:m31vels}: 
$38.3\pm2.7$~km~s$^{-1}$. This factor-of-two discrepancy (which 
implies a factor-of-four discrepancy between the true cluster mass and 
that estimated from our de-reddened total luminosity and 13-Gyr old 
population-synthesis $M/L_V$ ratio) could be resolved by
lowering the mass-to-light ratio, but reducing $\Upsilon_V^{\rm pop}$ by a factor 
of 4 would require imply an age for B037 of only a few Gyr, in conflict with the SED-fitting
results. Reducing the reddening of B037 to $E(B-V)\sim 0.85$ would yield a total
luminosity in better agreement with the measured $\sigma_v$, but this
would also conflict with the \citet{ma06b} SED-fitting. The only viable 
resolution appears to be the suggestion by \citet{cohen06} that the dust
lane projected across the face of the cluster \citep[reported by][]{ma06a} 
distorts measurements of extinction from integrated photometry.
This should be testable with color profiles and star-count analysis from
the ACS images. Such analysis is beyond the scope of the present paper,
and the properties of B037 computed using $E(B-V)=1.36$ are used in our
subsequent analysis.

\subsection{Quality of Fit for Different Models}
\label{subsec:chicomp}

\begin{figure}
\plotone{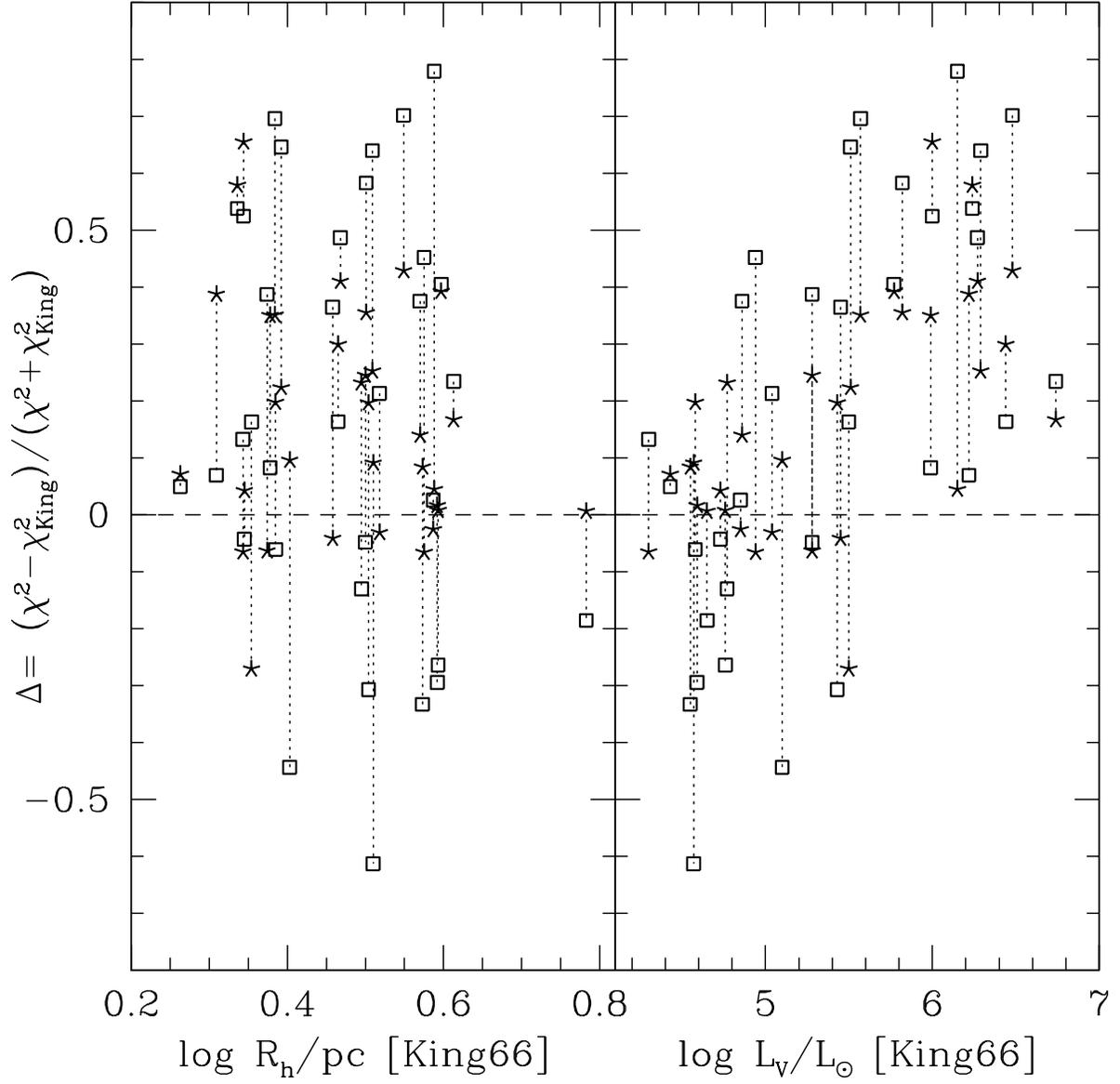}
\caption{
Comparison of goodness of M31 cluster model fits as a function of \citet{kin66} model 
size $R_h$ and luminosity $L_V$. Models are compared with 
$\Delta \equiv (\chi_{\rm alternate}^2-\chi_{\rm King}^2) / (\chi_{\rm alternate}^2+\chi_{\rm King}^2)$
which is positive if the \citet{kin66} model is a better fit.
Asterisks are \citet{wil75} models and open squares \citet{sersic} models.
\label{fig:chicomp}
}
\end{figure}

\citet{mv05} and \citet{dean06} give detailed discussions of the differences between various
cluster models and their applicability to GCs in galaxies other than M31.
Which structural model best represents the M31 clusters? Figure~\ref{fig:chicomp}
compares the ${\chi}^2$ values for the model fits as a function of cluster size
$R_h$ and total (model) luminosity $L_V$. The difference between models is parameterized
as 
\begin{equation}
\Delta \equiv (\chi_{\rm alternate}^2-\chi_{\rm King}^2) /
         (\chi_{\rm alternate}^2+\chi_{\rm King}^2)
\end{equation}
which is independent of the ${\chi}^2$ re-scaling; significantly better fits
of one model over another are signaled by $|\Delta|\gtrsim0.2$.
In general, \citet{kin66} models
fit the M31 cluster data better than \citet{wil75} or \citet{sersic} models. 
Cluster size does not strongly affect $\Delta$, but luminosity does:
\citet{kin66} models are more strongly preferred for more luminous M31 clusters. 
The preference for \citet{kin66} models for M31 clusters differs
from results using the same profile-fitting software for 
for clusters in the Milky Way, Magellanic Clouds, and Fornax dSph
\citep{mv05} and NGC~5128 clusters \citep{dean06}, where
\citet{wil75} models fit as well as or better than than \citet{kin66} models.
The results for M31 clusters do not appear to depend on
the image saturation for some clusters --- the clusters observed with ACS/HRC,
where saturation was not a problem, were also better-fit with \citet{kin66} models.
The fitting procedure returned higher background levels for M31 clusters than those in NGC~5128, but
the background value does not correlate with $\Delta$.

\begin{figure}
\plotone{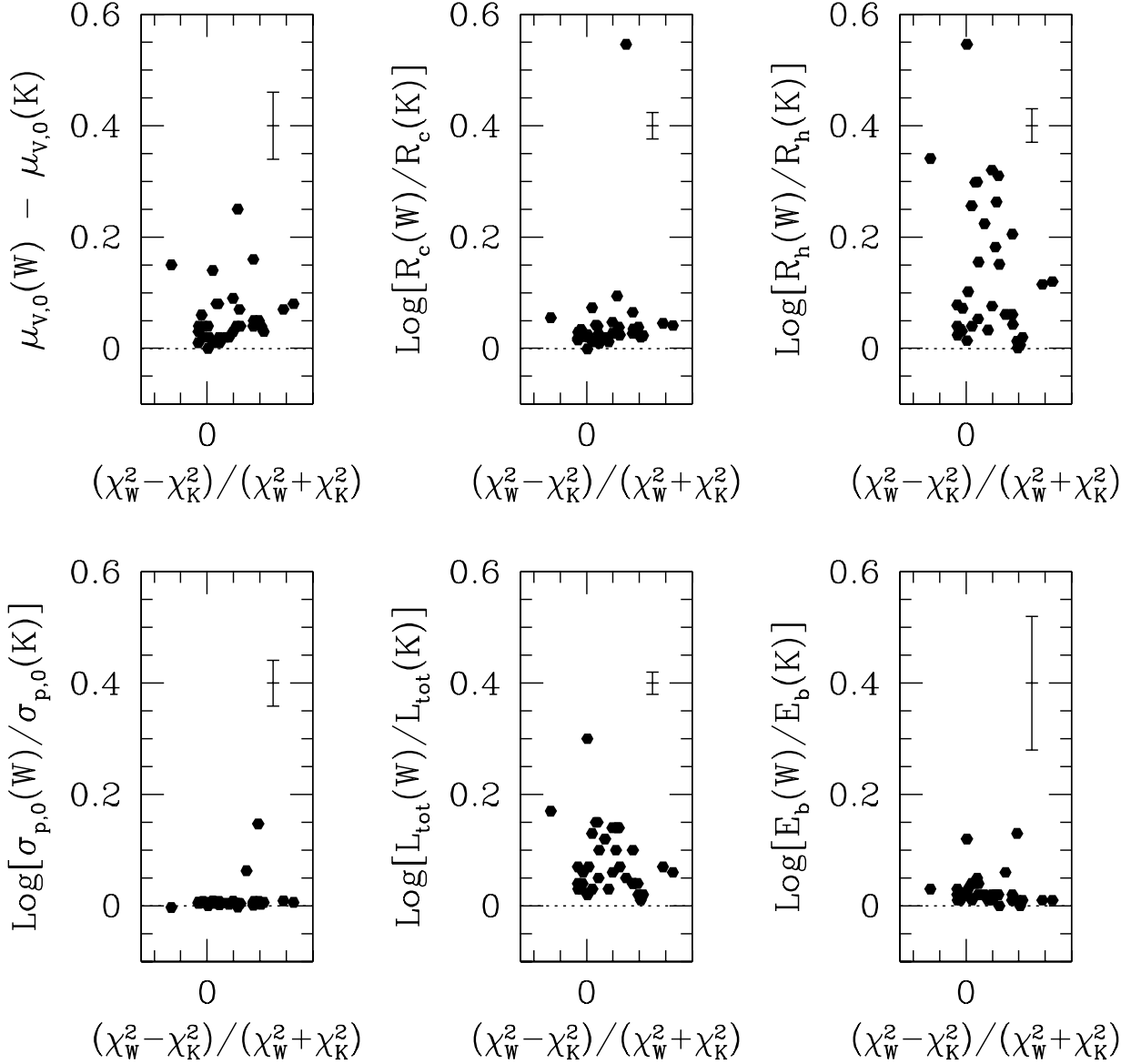}
\caption{
Comparison of \citet{wil75} and \citet{kin66} model-fit parameters for M31 
clusters as a function of goodness-of-fit difference $\Delta=
(\chi_{\rm Wilson}^2-\chi_{\rm King}^2) / (\chi_{\rm Wilson}^2+\chi_{\rm King}^2)$.
Errorbars show typical parameter uncertainties in a single model fit.
\label{fig:wils_comp}
}
\end{figure}

Figure~\ref{fig:wils_comp} compares the values of cluster
parameters derived from \citet{kin66} and \citet{wil75} model fits. 
The parameters do not clearly vary with goodness-of-fit,
and even the major outliers are not always much better fit with one model or the other. 
The mean offsets between parameters derived
for the same cluster from \citet{kin66} and \citet{wil75} models are
comparable to the rms scatter, and only slightly larger than the typical
parameter uncertainties: 
$\delta \mu_{V,0}=0.04\pm0.18$, $\delta(\log  R_h)=0.14\pm0.13$,
$\delta (\log L_V)=0.08\pm0.06$, and $\delta(\log R_c)=0.03\pm0.02$.
Small-scale systematic differences between the models (for example,
\citet{wil75} models, which have more extended halos, always return fainter central
surface brightnesses $\mu_{V,0}$ than \citet{kin66} models) do not affect
the global cluster parameters, such as the binding energy $E_b$ and
total luminosity $L_V$. These are very similar in all three of \citet{kin66}, \citet{wil75}
and \citet{sersic} models.
We suspect that the preference for \citet{kin66} over \citet{wil75} models is
due to some more subtle feature of the observational data that we have not yet been able to isolate, 
rather than an intrinsic difference
between M31 clusters and those in other galaxies. However, since a majority of the M31
clusters have parameters derived only with \citet{kin66} model fits, we use these
models as the basis for the fundamental plane analysis presented in \S\ref{subsec:fpcor}.
We do not expect the following discussion to depend on the model choice.

\section{Discussion}

We now combine the M31 GC parameters newly derived in the previous section
with those for clusters in the Milky Way and Magellanic Clouds \citep{mv05}
and NGC~5128 \citep{dean06} to form a large sample of GCs that have been
analyzed in a nearly homogeneous way. The sample comprises 
291 clusters in six galaxies, an unprecedentedly large sample for defining
the GC fundamental plane.
In the following analysis, cluster luminosities are derived from the model fits
to the surface photometry rather than integrated or aperture measurements.
For the M31 and NGC~5128 clusters, model and measured luminosities are found to have excellent
correspondence, while for Milky Way and Magellanic Cloud clusters, discrepancies 
between model and measured luminosities are often attributable to the measurement
aperture being smaller than the clusters' full size \citep{mv05}.
Where M31 clusters are compared with NGC~5128 clusters from \citet{dean06},
the 27 clusters omitted from that analysis are also omitted here.
As discussed above, \citet{kin66} model parameters are adopted for clusters
in all galaxies.

\subsection{Ellipticity distribution}

There are several possible explanations for the elongation of globular clusters:
\citet{lar01c} lists elongation as possibly resulting from  galaxy tides, 
internal rotation, cluster mergers \citep[crossing times in star cluster complexes
are short;][]{fk05}
and `remnant elongation' (due to accretion?) from some clusters' former lives as
dwarf galaxy nuclei.
The latter two mechanisms would seem to be applicable to only a small
fraction of clusters, probably the more massive one. They could be related to the
observations that the most luminous young and old
LMC clusters tend to be more flattened, and that 
the brightest globular clusters in both the Milky Way and M31 are also the most 
flattened \citep{vm84,vdb96a}.%
\footnote{As discussed in \S\ref{subsec:indiv}, there is some controversy about whether
G001 or B037 is the most luminous M31 GC, but both are quite elliptical.}
Measurement of GC ellipticities is an uncertain business particularly 
for faint clusters: the shape of 
outer isophotes is affected by the galaxy background light and low
signal-to-noise. For the extragalactic clusters measured with ELLIPSE,
very low values of ellipticity are disfavored because the ELLIPSE algorithm 
diverges at $\epsilon=0$. With these caveats in mind, it is still
worthwhile to explore ellipticity distributions and correlations with other 
parameters for clues to the origin of GC flattening.

\begin{figure}
\plotone{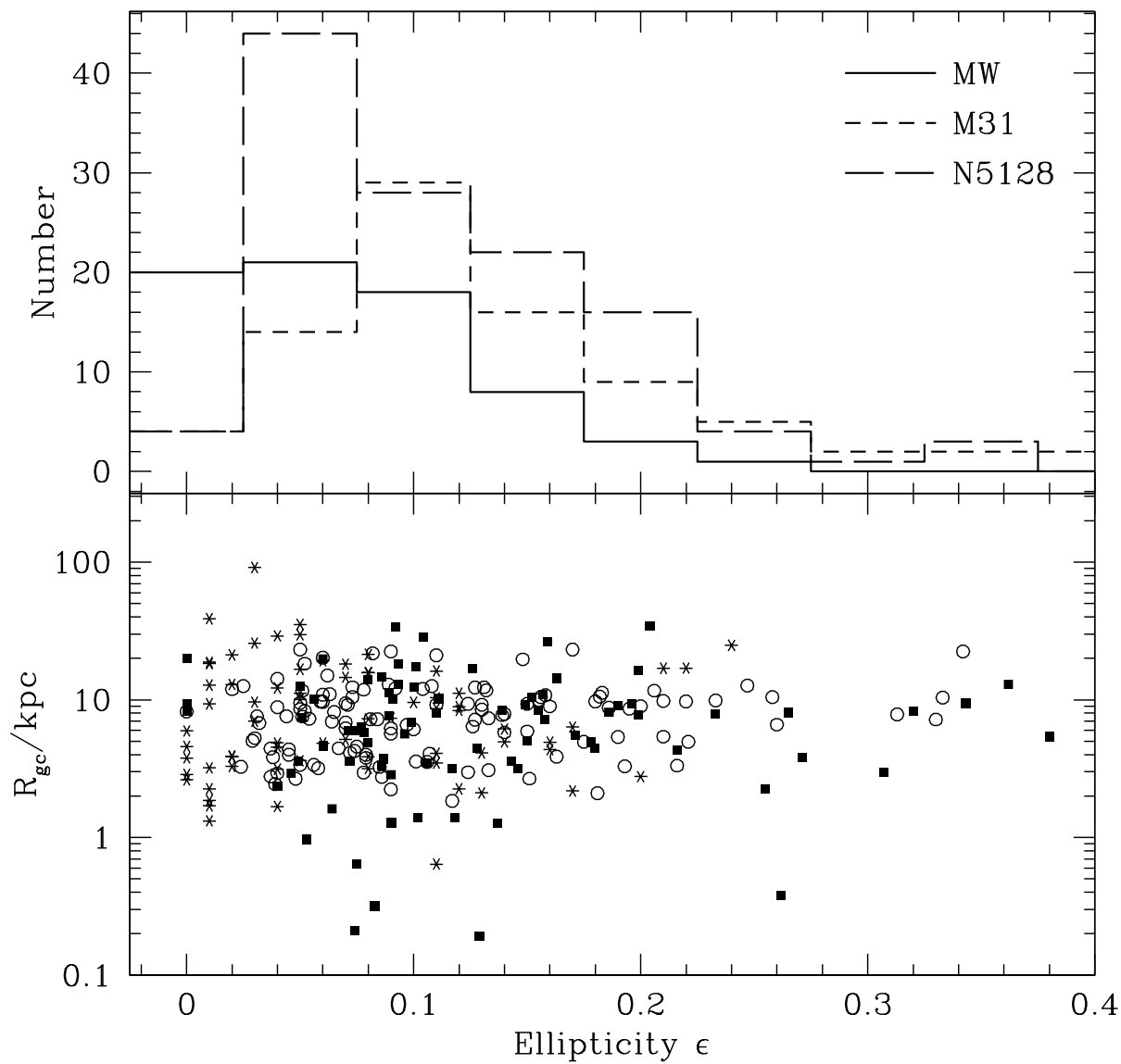}
\caption{Top: distribution of ellipticity for globular clusters in 
the Milky Way (solid line), M31 (short-dashed lines), and NGC~5128 (long-dashed lines).
Bottom: ellipticity versus projected galactocentric radius for globular clusters 
in M31 (filled squares), NGC~5128 (open circles), and the Milky Way (stars; 
three-dimensional $R_{\rm gc}$ used).
\label{fig:ellip_hist}
}
\end{figure}

Testing the idea that GC elongation is due to galaxy tides can be done to some
extent by comparing the ellipticities of clusters in different galaxies, and
thus different dynamical environments. Figure~\ref{fig:ellip_hist} shows the
distribution of ellipticities for clusters in the Milky Way \citep{h96}, M31 (this work), 
and NGC~5128 \citep{harris06}.
Compared to Figure~7 of \citet{hhhm02}, the number of M31 clusters is about 10\%
larger (although the sample is somewhat different and includes more very luminous clusters);
the number of NGC~5128 clusters is about 3 times larger and the distribution
of their ellipticities somewhat more skewed to lower values.
The distributions of ellipticities for M31 and NGC~5128 are not statistically
different; both differ from the Milky Way distribution in having few very round clusters
(likely an artifact of the measurement technique).
The bottom panel of the Figure shows the distribution of ellipticity with
galactocentric position: no correlation is evident.
From this comparison there is no clear evidence for major differences in ellipticity
distribution between the three different galaxies' GCs, and thus no evidence that the overall
galaxy environment is a major factor.

\begin{figure}
\plotone{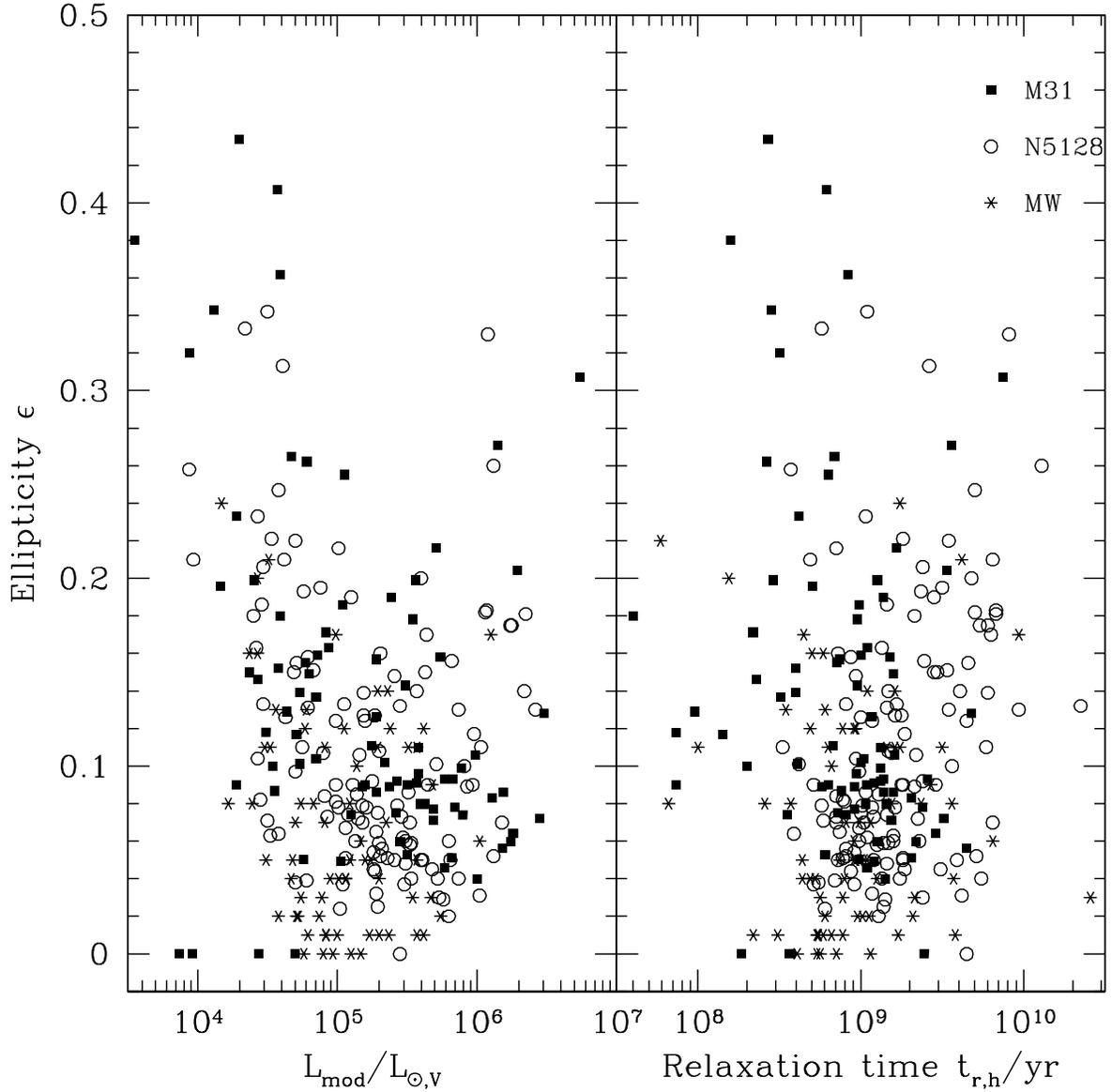}
\caption{Ellipticity versus luminosity and half-mass relaxation time for globular clusters 
in M31 (filled squares), NGC~5128 (open circles), and the Milky Way (stars).
\label{fig:mv_ellip}
}
\end{figure}

If cluster ellipticities are caused by internal processes such as rotation
or velocity anisotropy, then relaxation through dynamical evolution should act
to reduce any initial flattening \citep{fsk06}. More dynamically-evolved clusters might be 
expected to be rounder (although the relaxation time is long in the outer
regions of clusters which heavily influence measured ellipticity).
Figure~\ref{fig:mv_ellip} shows
ellipticity as a function of luminosity and half-mass relaxation time for clusters
in the three galaxies. 
A mild systematic decrease in $\epsilon$ with increased $L$ is visible, though with considerable
scatter; measurements of very large ellipticities for clusters with 
low luminosities are more likely to be spurious.
While the observation of \citet{vdb96a} that the most luminous clusters in M31 and the Milky Way 
are highly elliptical remains true, it is clear that all three galaxies also have very luminous clusters 
which are quite round: did these bright, round clusters simply have less angular momentum at formation?
Or are they the `true' globular clusters while the more flattened luminous clusters
are remnant dwarf galaxy nuclei somehow flattened while being removed from their original
galaxies? Further modeling of the dwarf galaxy stripping process might help to answer this question.
No correlation of $\epsilon$ with the more important dynamical quantity $t_{rh}$ is evident:
the data do not appear to favor the contention that more evolved clusters are rounder.
While primordial angular momentum still seems to be the most straightforward
explanation for GC ellipticity, the observed distribution appears to be due to 
a number of factors.

\subsection{Correlations with position and metallicity}
\label{subsec:posmet}

Structures of GCs in the Milky Way have been shown to be largely independent of
galactocentric distance and metallicity \citep{dm94,dj95,mcl00}, except for
the correlation of half-light radius with $R_{\rm gc}$ first noted by \citet{vmp91}.
Studies of cluster
structures in nearby galaxies \citep{bhh02,hhhm02} have found similar results, with the
exception of a mild correlation between cluster half-light radius and metallicity:
more metal-rich clusters tend to be smaller. The same effect is seen in more distant galaxies 
where the clusters' structures are barely resolved.
\citet{bs06} discuss several possible explanations and conclude that the $R_h-{\rm [Fe/H]}$
correlation is most likely a consequence of the correlation between $R_h$ and $R_{\rm gc}$ 
and the tendency of metal-rich GCs to be located at smaller $R_{\rm gc}$. 

\begin{figure}
\plotone{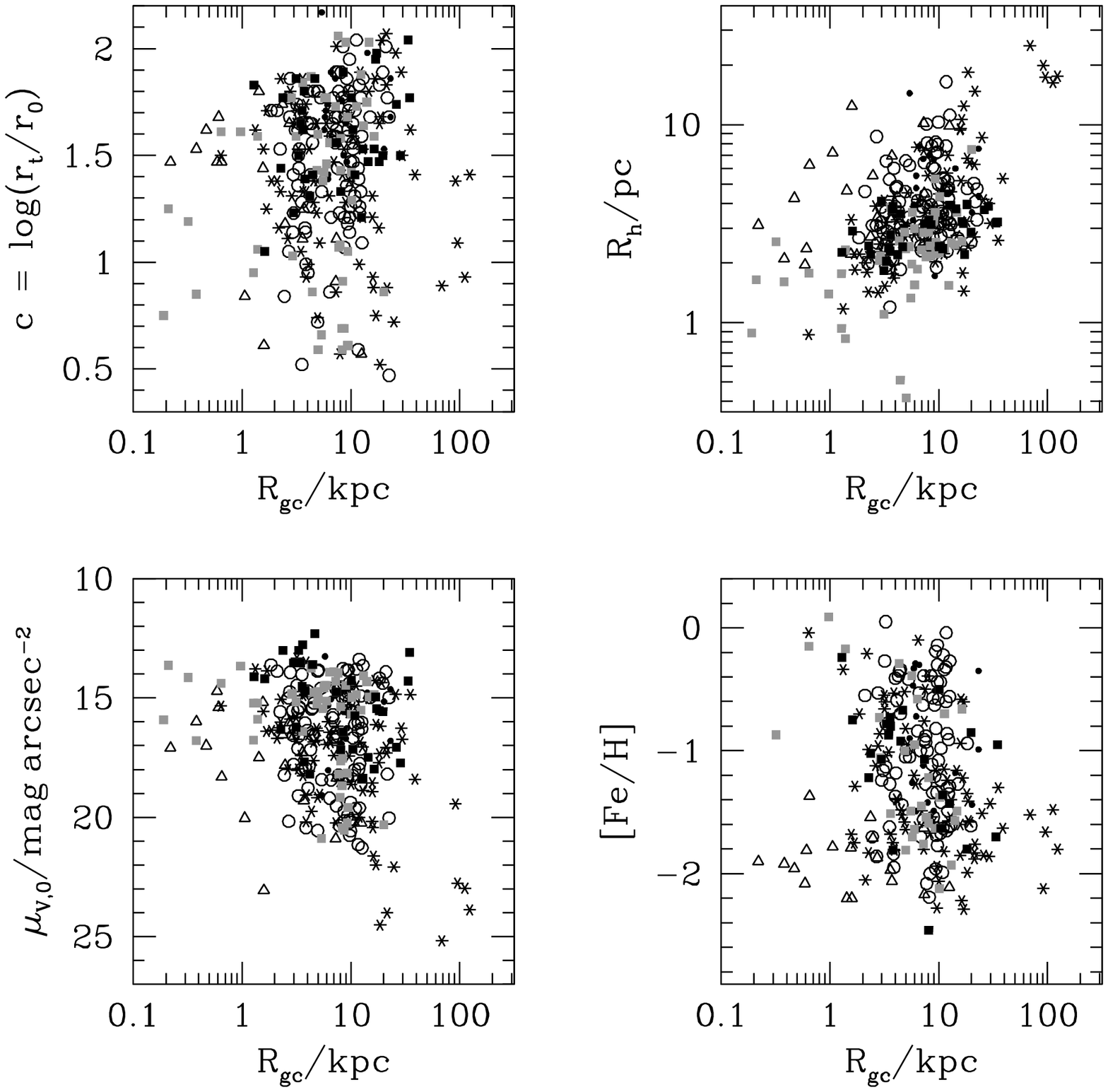}
\caption{Structural parameters as a function of galactocentric distance $R_{\rm gc}$
for globular clusters 
in the Milky Way (stars), the Magellanic Clouds and Fornax dwarf spheroidal (open triangles),
NGC~5128 [open circles are clusters from \citealt{dean06}; small filled circles 
are from \citealt{hhhm02}], and M31 (filled squares). The darker filled squares are the 
M31 clusters observed with ACS or STIS; lighter squares are WFPC2 observations.
$R_{\rm gc}$ is three-dimensional for Milky Way clusters but projected two-dimensional
for other galaxies.
\label{fig:rgc_struct}
}
\end{figure}

Figure~\ref{fig:rgc_struct} plots structural parameters and [Fe/H] for our sample as
a function of galactocentric distance $R_{\rm gc}$. For the Milky Way clusters
the true three-dimensional distance is used, while for external galaxies only
projected distances are available. For the M31 clusters, no correlation of 
concentration $c$ with $R_{\rm gc}$ is seen.  Central surface brightness $\mu_{V,0}$ 
becomes fainter with increasing $R_{\rm gc}$ but this is likely to be a selection
effect: fainter clusters are difficult to see against the bright background of
the M31 bulge (\citealt{dean07} discuss the same effect in detail for NGC~5128).
As expected, $R_h$ increases with $R_{\rm gc}$, although the
trend is largely driven by the innermost clusters in M31. In the Milky Way
the $R_h-R_{\rm gc}$ trend is driven both by small inner clusters and the large
clusters at very large distances from the Galactic center. Such distant, low-surface
brightness clusters will be difficult to detect in external galaxies, and their
absence from the present sample may well be due to selection effects.
For the clusters in NGC~5128, which cover a more restricted range in $R_{\rm gc}$, 
\citet{dean07} conclude that the $R_h-R_{\rm gc}$ trend is weak and only
marginally significant. For the M31 and Milky Way clusters, a weak correlation
of [Fe/H] with $R_{\rm gc}$ is present, but again is driven mainly by
clusters at the extremes of $R_{\rm gc}$.
The Magellanic Cloud and Fornax dSph clusters are offset from the larger
galaxies in all of these plots, as might be expected from the parent galaxies' smaller sizes.
Scaling cluster properties by galactocentric distance as in \citet{mcl00}
is therefore expected to be useful in computing the `fundamental plane'
for these clusters.

There are half a dozen very small ($R_h \lesssim 1$~pc) M31 clusters visible in the 
upper right panel of Figure~\ref{fig:rgc_struct}. Such clusters are not found in 
the other galaxies studied, which makes their presence in M31 slightly suspicious.
Of the six clusters, only the largest, B155, has a confirmatory radial velocity
\citep{gal06} while the remainder are classified by those authors as ``cluster candidates''.
The compactness of these objects makes assessing their nature problematic, even
with the use of {\it HST} imaging. They are a reminder that the structural analysis of
star clusters in distant galaxies will always be problematic: existing catalogs are
neither fully complete nor entirely reliable.

\begin{figure}
\plotone{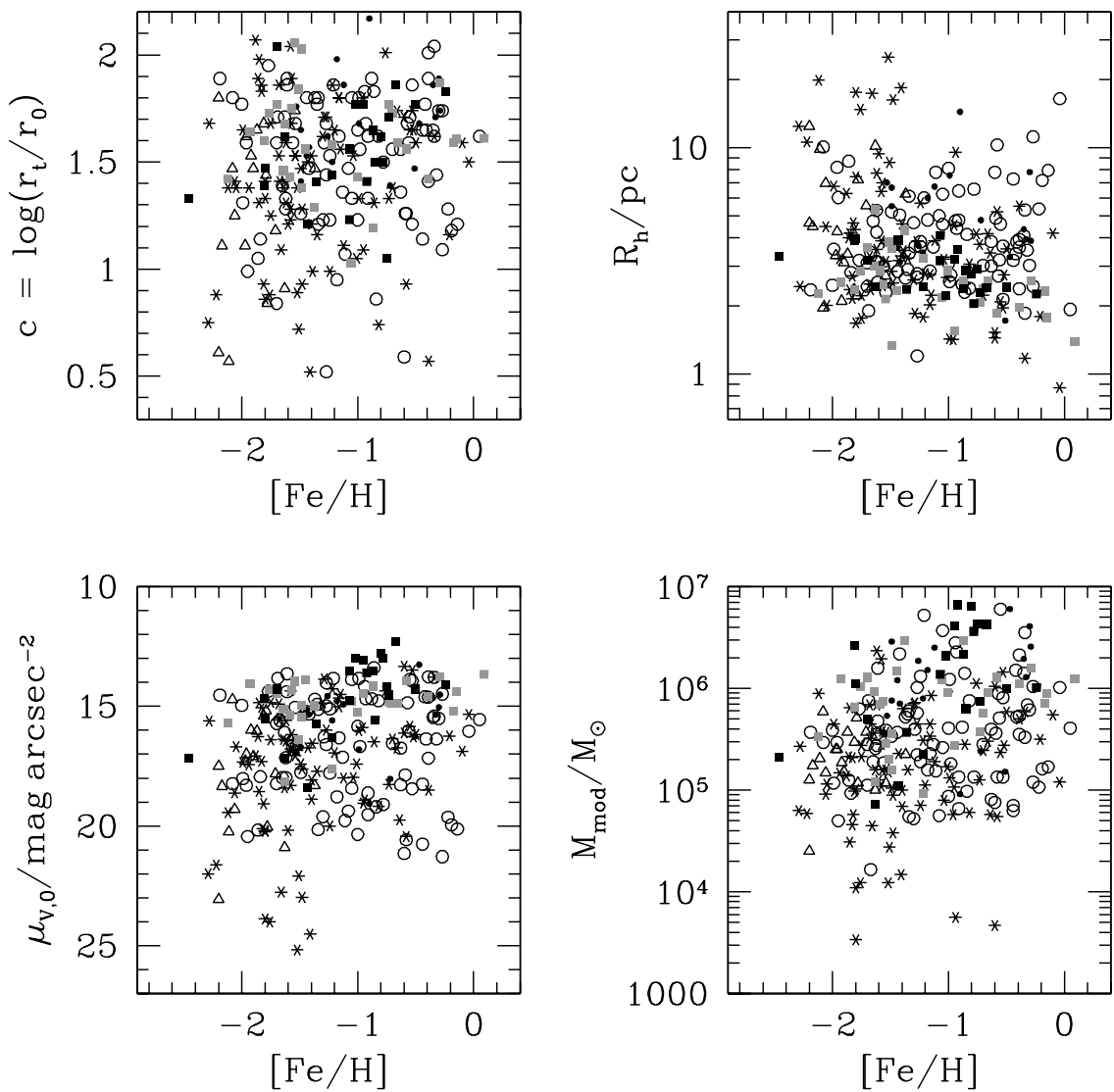}
\caption{Structural parameters as a function of metallicity for globular clusters.
Symbols as in Figure~\ref{fig:rgc_struct}.
Only clusters with measured values of [Fe/H] (not assumed average values) are shown. 
\label{fig:feh_struct}
}
\end{figure}

Figure~\ref{fig:feh_struct} plots cluster
structural parameters as a function of [Fe/H]; only clusters with measured values of
[Fe/H] (not assumed average values) are shown. 
No correlation of concentration $c$  with [Fe/H] is seen.
A weak correlation of $R_h$ with [Fe/H] is present in the expected sense:
$R_h$ decreases with increased metallicity. 
The exception is NGC~5128, for which the best-fit slope of
$\log R_h$ against [Fe/H] is zero; the present sample of clusters in this galaxy also has
a lower metallicity gradient ([Fe/H] vs.\ $R_{\rm gc}$) than in other galaxies.
The $R_h-$[Fe/H] corrlation weakens if only clusters with $2<R_{\rm gc}<20$~kpc are considered, 
suggesting that a different mix of bulge and halo clusters in M31
and the Milky Way compared to NGC~5128 may be responsible.
For central surface brightness $\mu_{V,0}$ there is a slight
systematic increase with [Fe/H], likely because of the weak correlation of [Fe/H] 
and $R_{\rm gc}$. Finally, the lower-right panel shows that there is
a weak mass-metallicity correlation: this is due entirely to our use of population
synthesis model mass-to-light ratios, which increase rapidly above
${\rm [Fe/H]}=-1$ \citep[see Figure~1 of][]{dean07}.
All of these trends,
however, are mild compared with the correlations with cluster luminosity
to be presented below.

Figure~\ref{fig:lum_struct} shows some of the correlations between cluster parameters
and luminosity, for GCs in M31, the Milky Way, NGC~5128, the Magellanic Clouds (MCs), and
the Fornax dSph. The properties of clusters in all six galaxies fall in the same
regions of parameter space. The selection effects discussed above are apparent
in that there are fewer faint and low-surface-brightness clusters in the extragalactic samples.
The new M31 ACS/STIS observations in particular
do not contain any faint, low-concentration clusters, partly because the ACS
observations targeted luminous M31 clusters and also because very low-concentration
M31 clusters are too well-resolved to be analyzed with the ellipse-fitting method used
(see \S~\ref{subsec:SBprofs}). There are few very large ($R_h>10$~pc) extragalactic
clusters, probably because these are difficult to recognize and more easily confused 
with background galaxies. The right-hand panels of the Figure show the lack of 
correlation between $R_h$ and $L$ and the strong correlation of $E_b$ with $L$ which
are both hallmarks of the globular cluster fundamental plane. The fundamental plane
correlations evidently hold over nearly 4 decades in luminosity and in a variety
of environments. 

\subsection{The Fundamental Plane}
\label{subsec:fpcor}

Correlations between globular cluster structural parameters are 
often summarized as describing a ``fundamental plane'' analogous to (but different from) 
that for elliptical galaxies. There are several equivalent formulations of the globular cluster 
fundamental plane in the literature \citep{bur97,dj95,mcl00,dean07}.
Here we concentrate on the relationships between half-light radius, mass, and binding energy
to search for differences between the fundamental planes of clusters
in different galaxies. An important caveat to this analysis is that the GC masses used are
derived from population synthesis model-predicted mass-to-light
ratios, not from directly observed velocity dispersions (see \S~\ref{subsec:apvel}).

One of the many ways of looking at the globular cluster fundamental plane
is a plot of projected cluster half-light radii as a function of mass.
Modulo some stretching, this plot is equivalent to one of projected half-light 
surface density $\Sigma_h$ versus $M$, which in turn is the largest part of
the $\kappa_{m,1}$ and $\kappa_{m,2}$ parameter space. A different reformulation
makes $R_h$ vs.\ $M$ equivalent to binding energy $E_b$ vs.\ $M$:
$E_b = f(c) M^2/R_h$ where $f(c)$ is a weak function of the cluster concentration
\citep[and see below]{mcl00}.
Figure~\ref{fig:rh_m} shows $R_h$ versus $M$ for GCs in all of our galaxies. As 
is well known from previous investigations of the Galactic GC system
\citep{vmp91, dm94}, there is no clear, monotonic mass-radius relation for globular 
clusters. However, at masses $M\ga 1.5\times10^6\,M_\odot$ there is an 
apparent upturn in the lower envelope of $R_h$ versus $M$. \citet{dean07}
interpret this increasing lower bound on $R_h$ for 
clusters as an extension of a very similar relation for early-type 
galaxies. The latter relation follows from the existence of a ``zone of 
exclusion'' in the (luminosity-based) $\kappa$ space of \citet{bur97}, 
which can be interpreted as a mass-dependent {\it upper} limit 
on the average surface density $\Sigma_h$. The bound $R_h^{\rm min}(M) 
= [M/2\pi\Sigma_h^{\rm max}]^{1/2}$ for galaxies, simply extrapolated 
down into the GC mass range, is drawn as the dash-dot line in Figure~\ref{fig:rh_m}; 
see \citep{dean07} for further details.

The two dashed lines in the lower-left corner of the Figure
represent a different bound on GC size: evaporation by two-body relaxation, 
which is expected to occur within 20--40 initial half-mass relaxation times
\citep[e.g.][]{go97}. The two lines correspond to $t_{\rm rh}=(13/20)$~Gyr (upper)
and $t_{\rm rh}=(13/40)$~Gyr (lower); clusters which formed with masses and sizes 
to the lower-left of these lines are expected to have evaporated by the present day.
Several of the M31 clusters are below the `evaporation' lines, indicating that they 
must have evolved into this region of the diagram and may in fact be dissolving. 
The most extreme of these are the very compact M31 clusters discussed in 
\S~\ref{subsec:posmet}. The identfication of these objects as M31 GCs is 
somewhat suspect: high-resolution spectroscopic observations would be of interest
to confirm their status and measure velocity dispersions.

None of the galaxies' clusters show evidence for correlation of half-light radius 
with mass, but there are differences between galaxies in both position and scatter 
of their clusters in Figure~\ref{fig:rh_m}.
The larger scatter for Milky Way clusters compared to those in M31 and NGC~5128
can be attributed to the presence of more very diffuse clusters in the MW sample
\citep[such clusters are difficult to detect in the extragalactic systems;][]{dean07}
and greater distance uncertainties for individual Galactic GCs.
The M31 sample has very few diffuse clusters due to the selection effects described above,
and so is offset to smaller average $R_h$ than the Milky Way or NGC~5128 samples
The Magellanic Cloud and Fornax cluster samples are complete, however, and
have larger average $R_h$ than the large-galaxy GCs.

Is the difference between cluster sizes in different galaxies due to local environment?
The structure of tidally-limited clusters should depend on the local galaxy density, 
and cluster sizes are known to be correlated with galactocentric position (see the
upper right panel of Figure~\ref{fig:rgc_struct}). A better indicator of galactic
potential than galactocentric distance alone is:
\begin{equation}
R_{\rm gc}^* \equiv \frac{R_{gc}/(8\ {\rm kpc})}{V_c/(220\ {\rm km\ s}^{-1})}
\end{equation}
where $V_c$ is the galaxy circular velocity.%
\footnote{The normalization of $R_{\rm gc}^*$ ensures that quantities derived from it are 
equivalent to those plotted for Milky Way clusters in \citet{mcl00}.
It should properly be computed with the three-dimensional galactocentric
distance, but only the projected distance is available for the extragalactic clusters.
}
We use values $V_c$ in km~s$^{-1}$ of: Milky Way, 220; 
M31, 230 \citep{car06}; NGC~5128, 190 \citep{wood06,peng04a,peng04b,mat96};
LMC, 65 \citep{vdm02}; SMC, 60 \citep{stan04}, and Fornax dSph, 18 \citep{walk06}.
For a galaxy whose total mass distribution
is an isothermal sphere, mass density $\rho \propto (V_c/R_{\rm gc})^2$, so
$R_{\rm gc}^* \propto {\rho}^{-1/2}$ is a guide to the local galaxy density. 
Figure~\ref{fig:rh_rgcstar} shows $R_h$ as a function of re-normalized galactocentric
distance $R_{\rm gc}^*$, with the four galaxies plotted separately.
The normalization has removed the $R_h/R_{\rm gc}$ offset between the Magellanic Cloud/Fornax clusters
and those in the large galaxies, indicating that the larger cluster sizes
in the small galaxies can be understood as due to the lower galactic densities.
The least-squares fit of $\log R_h = C + \gamma \log(R_{\rm gc}^*)$ is shown
for each galaxy, and given in Table~\ref{tab:fp_fits}. As might be expected,
there is less dependence of $R_h$ on $R_{\rm gc}^*$ for the extragalactic clusters
compared to the Milky Way GCs, due to the use of projected galactocentric distances in 
computing $R_{\rm gc}^*$.

The normalization by galactic potential should allow us to fairly
compare the fundamental planes of clusters in the different galaxies.
In analogy with \citet{mcl00} and \citet{hhhm02} we define the re-normalized quantities
\begin{equation}
R_h^* \equiv R_h\times (R_{gc}^*)^{-\gamma}
\end{equation}
and
\begin{equation}
E_b^{*} \equiv E_b \times (R_{gc}^*)^{\gamma}
\end{equation}
where $\gamma$ is derived from the fits given above.
Figure~\ref{fig:rhs_feh} shows that there is no correlation of $R_h^*$ with [Fe/H],
so any differences between galaxies are not due to residual effects of metallicity 
on the fundamental plane.

The top panel of Figure~\ref{fig:ebm_delta} shows $E_b^{*}$ versus $M$, with the
the corresponding least-squares fits for each galaxy given in Table~\ref{tab:fp_fits}.
The basic proportionality $E_b\propto M^2$ is built into the definition of 
$E_b$, and clusters in all galaxies follow this relation very tightly.
In particular, the fits for 
NGC~5128 and Milky Way clusters are identical within their errors, confirming
the \citet{hhhm02} result with a sample three times
larger and extending to much lower luminosities. 
At the high end of the mass range, the clusters in both M31 and NGC~5128 fall below
the best-fit lines, indicative of the `break' in
cluster properties at $\sim 1.5\times 10^6$~M$_\odot$ noted by \citet{dean07}. 
Below this break, the fits for all galaxies are similar, with the largest
differences at the low-mass end where the number of clusters is small
and measurement errors are large. 
The bottom panel of Figure~\ref{fig:ebm_delta} shows the difference between
$E_b^{*}$ predicted from the Milky Way fit and the measured values. 
Table~\ref{tab:fp_fits} gives the mean offsets for each galaxy.
The NGC~5128 and Magellanic Cloud clusters (and the Milky Way clusters, by
construction) have offsets consistent with zero, while the M31 clusters have
a slight offset which can be attributed to sampling issues: the lack
of diffuse clusters and the presence of the low-$R_h$ objects at $M\sim5\times10^4$~M$_\odot$

Globular clusters have a strikingly similar nature 
in the widely different environments we have sampled, including a dwarf
spheroidal, irregulars, large disk galaxies, and a giant elliptical.
The homogeneous and internally precise data we can now work with,
in addition to the much larger sample sizes of clusters, have 
established the trend of the fundamental plane securely, showing
that globular clusters in all these environments follow the first-order
trend $E_b \sim M^2$ quite accurately.  The scatter around this
basic relation is surprisingly small, showing that the structural
properties of these star clusters are far simpler than we could
have expected from theoretical arguments alone. Even the 
most massive GCs seem to show a fairly smooth transition between
the properties of clusters and those of dwarf galaxies.

\section{Summary}

New {\it Hubble Space Telescope} observations are used to construct surface brightness profiles for
34 M31 globular clusters, including the most massive clusters in the galaxy. Structural
models are fit to the surface brightness profiles: for reasons not yet understood,
M31 clusters are unlike those in other nearby galaxies in slightly preferring \citet{kin66} models 
over the more extended \citet{wil75} models. When combined with previously-published data,
we now have a comprehensive and publicly available compilation of a complete
suite of structural and dynamical parameters for as many GCs in M31 as in the Milky Way.
The structural parameters of
clusters in M31 and other local galaxies define a very tight fundamental plane with similar slopes
$E_b \sim M^2$,
showing the essential similarity of clusters in different environments over a range
of almost 4 decades in luminosity or mass. 
M31 clusters are slightly offset from the $E_b-M$ relation
defined by the Milky Way and NGC~5128, likely because of sampling effects.
In both M31 and NGC~5128 clusters, the most massive GCs trend towards
dwarf galaxies in the binding-energy/mass relation at about $10^6$~M$_{\odot}$.
The detailed modeling made possible by high-resolution imaging shows that
the overall properties of globular clusters in different galaxies are 
remarkably similar; the subtle differences noted here may point to differences
in the histories or environments of their parent galaxies.
The present dataset pushes our assessment of the fundamental plane
beyond the previous state of the field \citep{mcl00,hhhm02,bhh02,mv05},
making it possible to search for second-order trends in the FP line 
and probe the transition between the structures of globular clusters and galaxies.

\acknowledgments
We thank the referee for a rapid and thoughtful report.
Support to P.B. and D.E.M. for program GO-10260 was provided by NASA through a grant from the Space
Telescope Science Institute, which is operated by the Association of Universities
for Research in Astronomy, Inc., under NASA contract NAS 5-26555.
W.E.H. and G.L.H.H. thank the Natural Sciences and Engineering Research Council of Canada for 
financial support. D.A.F. thanks the Australian Research Council for financial support. 


\clearpage

\clearpage

\clearpage

\begin{figure}
\plotone{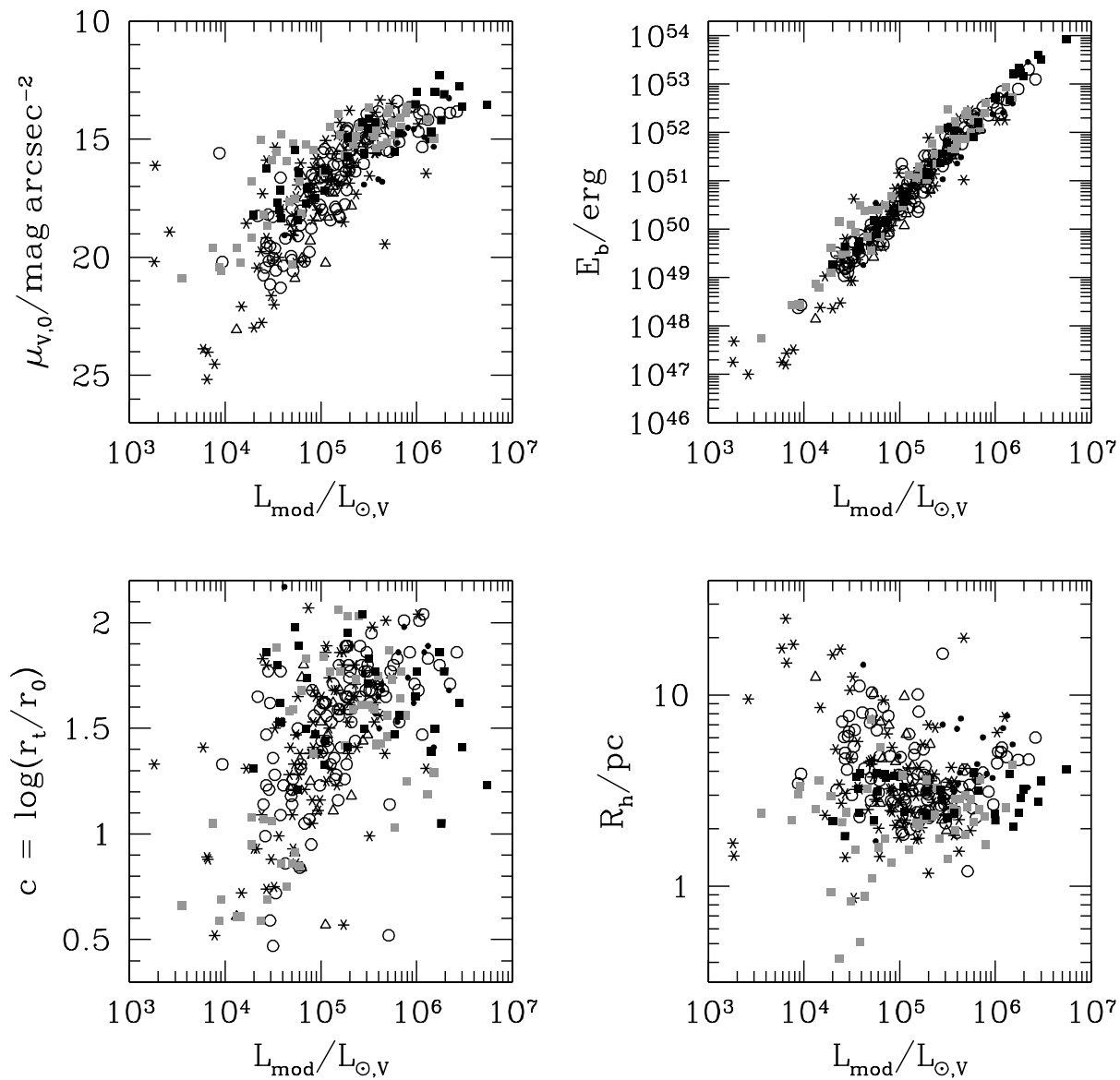}
\caption{Structural parameters as a function of luminosity for globular clusters.
Symbols as in Figure~\ref{fig:rgc_struct}.
\label{fig:lum_struct}
}
\end{figure}

\begin{figure}
\plotone{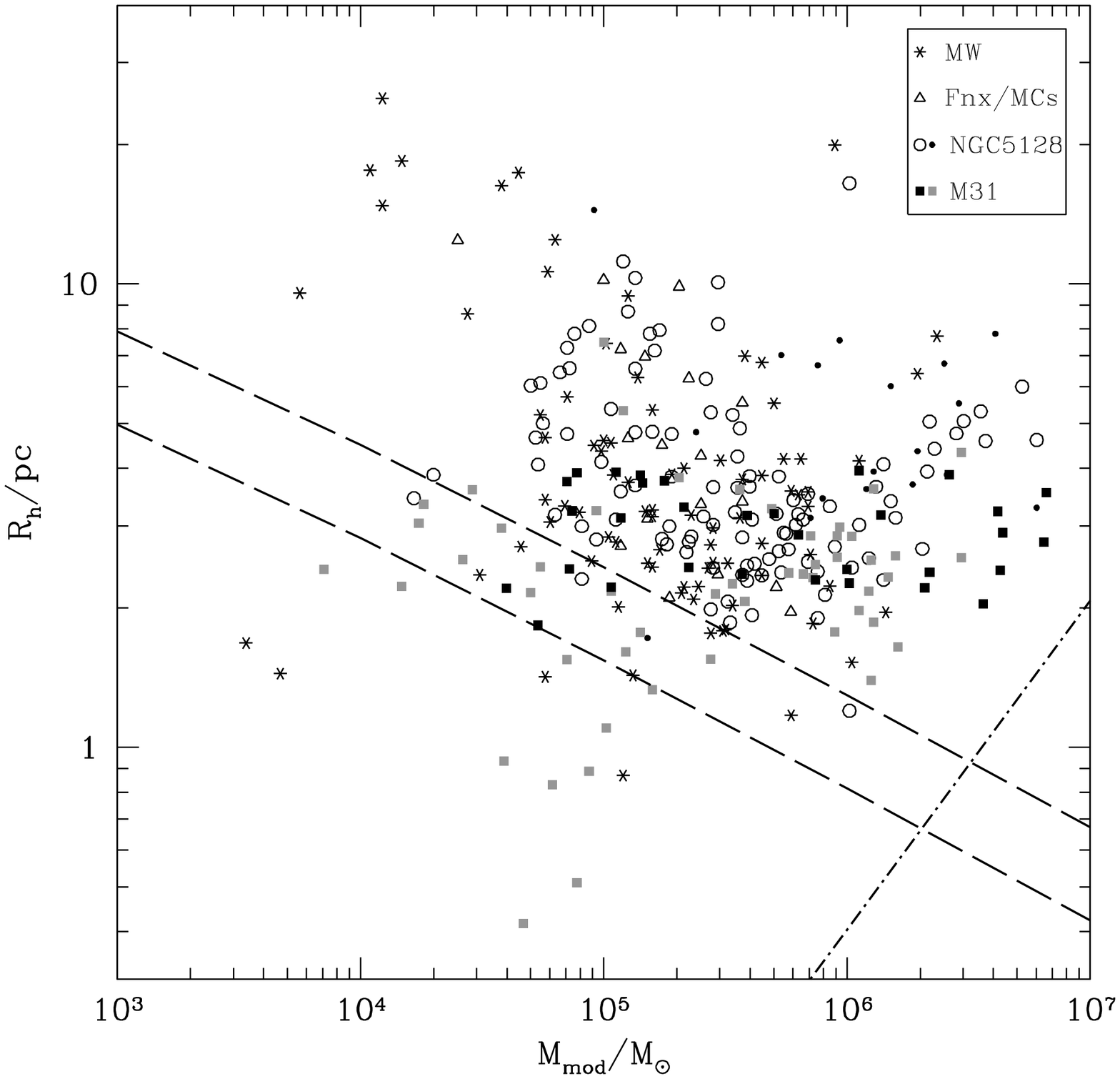}
\caption{Projected half-light radius as a function of cluster mass for
GCs in the Milky Way, Magellanic Clouds and Fornax, M31, and NGC~5128. 
Symbols: Milky Way (stars), the Magellanic Clouds and Fornax dwarf spheroidal (open triangles),
NGC~5128 [open circles are clusters from \citealt{dean06}; small filled circles 
are from \citealt{hhhm02}], and M31 (filled squares). The darker filled squares are the 
M31 clusters observed with ACS or STIS; lighter squares are WFPC2 observations.
Long-dashed lines are `survival' lines; dot-dashed line is lower $R_h$ bound 
defined by \citet{dean07}.
\label{fig:rh_m}
}
\end{figure}

\begin{figure}
\plotone{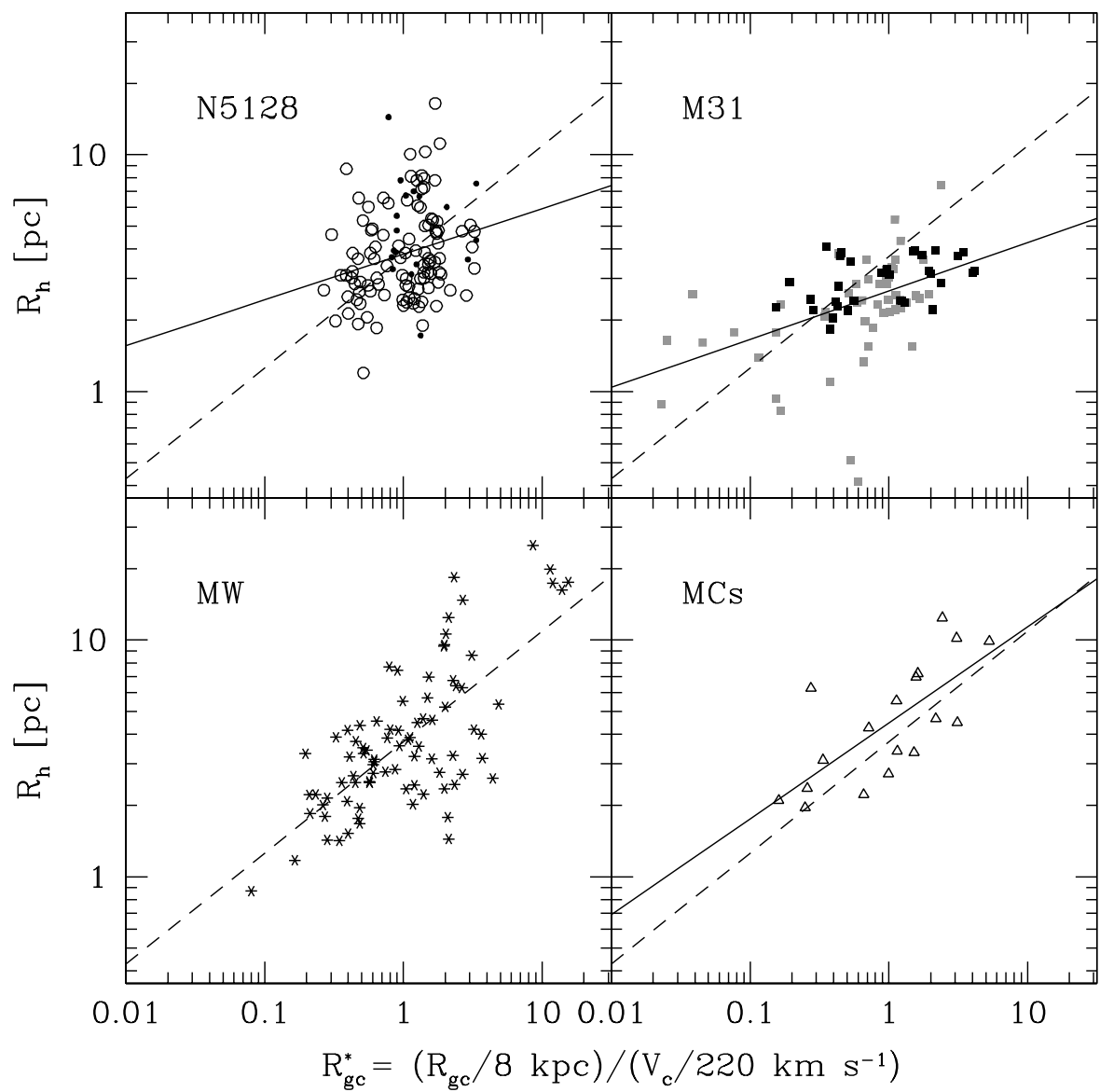}
\caption{Projected half-light radius 
as a function of re-normalized galactocentric distance $R_{\rm gc}^*$ for globular clusters in the Milky Way 
(lower left), Magellanic Clouds and Fornax (lower right), M31 (upper right), and NGC~5128 (upper left).
Solid lines are least-squares fits; dashed line in each panel is the Milky Way least-squares fit.
\label{fig:rh_rgcstar}
}
\end{figure}

\begin{figure}
\plotone{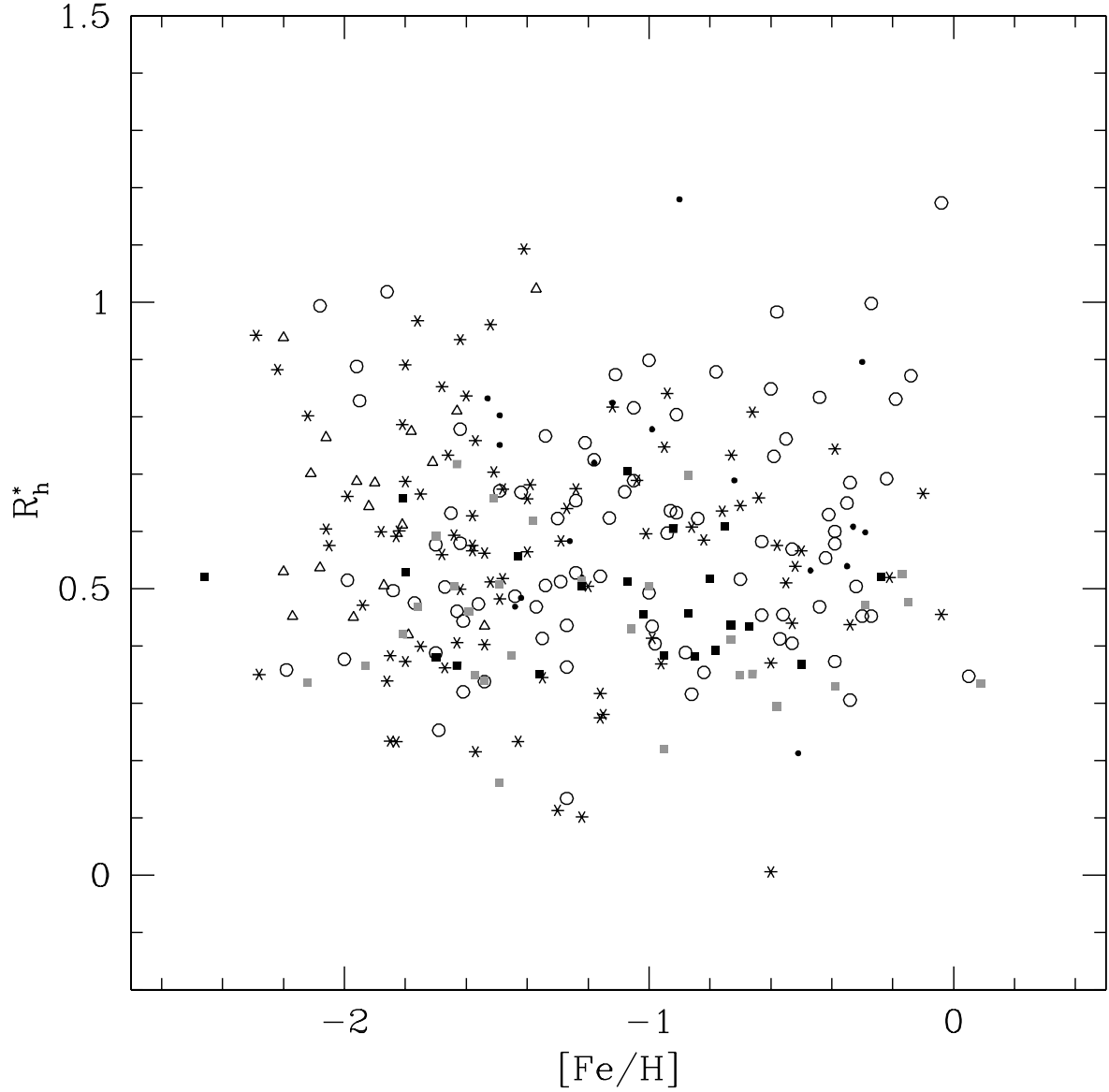}
\caption{Re-normalized half-light radius $R_h^*$ as a function
of  metallicity for globular clusters in the Milky Way, 
Magellanic Clouds and Fornax, M31, and NGC~5128. Symbols as in Figure~\ref{fig:rh_m}.
Only clusters with measured values of [Fe/H] (not assumed average values) are shown. 
\label{fig:rhs_feh}
}
\end{figure}

\begin{figure}
\plotone{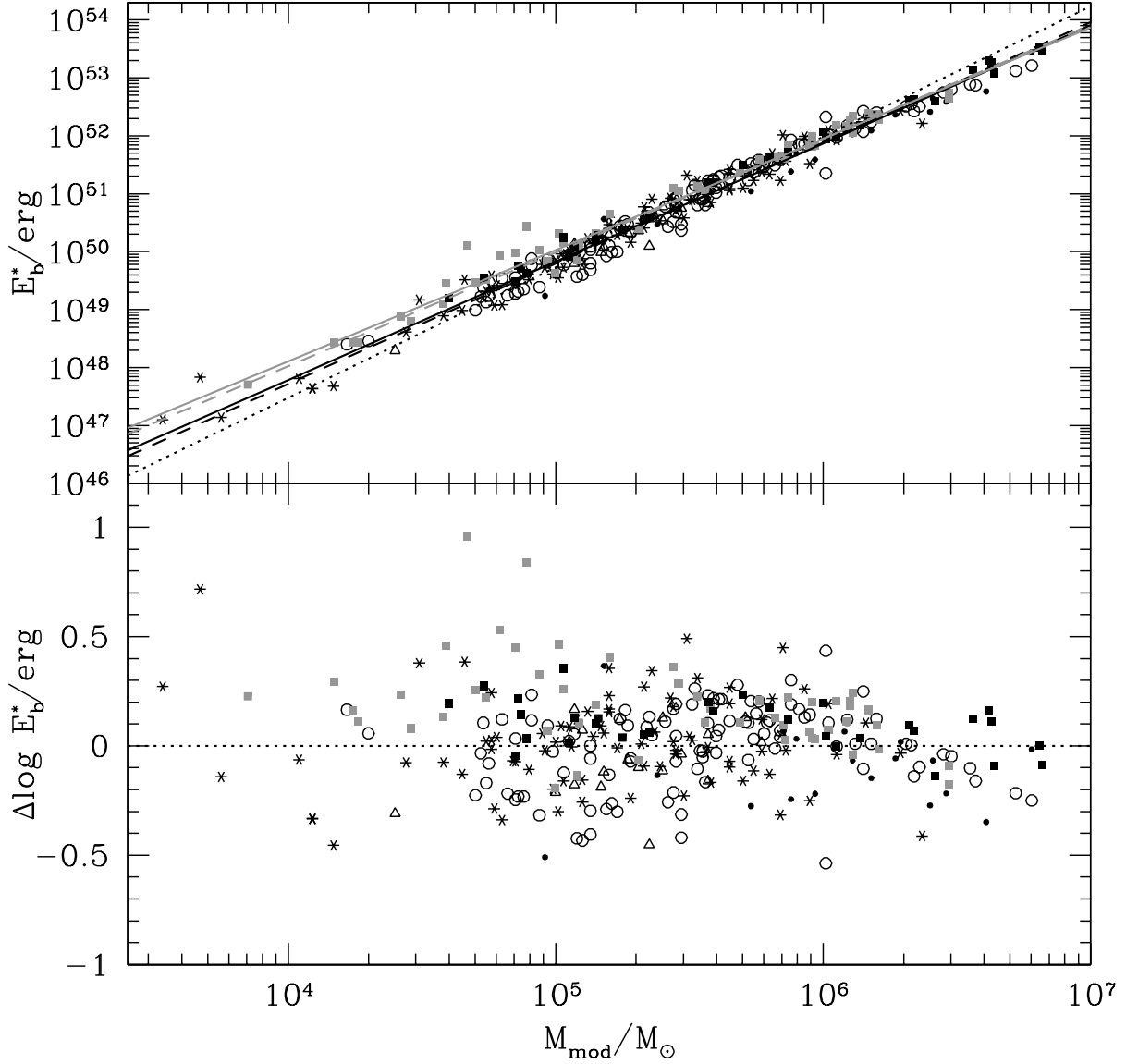}
\caption{Re-normalized binding energy $E_b^{*}$ (top) and difference from Milky Way
least-squares fit (bottom), as a function of mass for globular clusters in the Milky Way, M31, 
and NGC~5128.
Symbols as in Figure~\ref{fig:rh_m}; lines in top panel are least-squares fits
for Milky Way (black solid line), all M31 clusters (grey solid line), 
M31 ACS/STIS clusters only (grey dashed line), NGC~5128 clusters (black dashed line), 
and Magellanic Cloud and Fornax clusters (dotted line). 
\label{fig:ebm_delta}
}
\end{figure}

\begin{deluxetable}{lrrrrrrrl}
\tabletypesize{\scriptsize}
\tablecaption{Integrated measurements for M31 clusters\label{tab:m31meas}}
\tablecolumns{9}
\tablewidth{0pt}
\tablehead{
\colhead{Cluster} & \colhead{$\epsilon$} & \colhead{$\theta$} & 
\colhead{$V$ or alt.\tablenotemark{a}} & \colhead{$F814W$} & 
\colhead{$E(B-V)$}& \colhead{$R_{\rm gc}$} & \colhead{[Fe/H]} & \colhead{source\tablenotemark{b}}\\
\colhead{} & \colhead{} & \colhead{[\arcdeg E of N]} & 
\colhead{[VEGAMAG]} & \colhead{[VEGAMAG]} & 
\colhead{} & \colhead{[kpc]} & \colhead{} & \colhead{}
}
\startdata
\multicolumn{9}{l}{ACS/WFC targets from program 10260}\\
B037& 0.27& $ 54$ & 16.06 & 14.07 & 1.38   & 2.97 & $-1.07\pm0.2$ & P02 \\
B042& 0.10& $134$ & 15.71 & 14.29 & 0.77   & 3.31 & $-0.78\pm0.2$ & P02 \\
B063& 0.10& $ 28$ & 15.31 & 14.13 & 0.44   & 3.47 & $-0.87\pm0.3$ & HBK \\
B082& 0.05& $161$ & 15.02 & 13.63 & 0.72   & 3.61 & $-0.80\pm0.2$ & P02 \\
B088& 0.28& $178$ & 15.04 & 13.90 & 0.48   & 3.79 & $-1.81\pm0.1$ & P02 \\
B151& 0.07& $  8$ & 13.26 & 13.26 & 0.38   & 1.61 & $-0.75\pm0.2$ & HBK \\
\multicolumn{9}{l}{Serendipitous clusters observed in program 10260}\\
B056D& 0.22& $117$ &  18.81 & 17.72 & 0.08      & 4.20 & $-1.20\pm0.6$ & AVE \\
B041 & 0.05& $ 30$ &  18.95 & 18.13 & 0.13      & 2.72 & $-1.22\pm0.2$ & P02 \\
B061 & 0.14& $  2$ &  16.83 & 15.62 & 0.37      & 3.58 & $-0.73\pm0.3$ & P02 \\
B090 & 0.14& $124$ &  18.49 & 17.52 & 0.08      & 3.15 & $-1.20\pm0.6$ & AVE \\
B147 & 0.10& $172$ &  16.03 & 15.04 & 0.08      & 1.29 & $-0.24\pm0.4$ & HBK \\
B162\tablenotemark{c}  
     & 0.30& $ 47$  & 17.62 & 16.54 & 0.31      & 2.27 & $-1.22\pm0.2$ & B00\\
\multicolumn{9}{l}{ACS/HRC targets from programs 9719 and 9767}\\
B023 & 0.13& $ 60$  & 13.88 & 12.72 & 0.36      &  4.45& $-0.92\pm0.1$ & HBK \\
B158 & 0.05& $ 13$  & 14.36 & 13.44 & 0.12      & 2.38 & $-1.02\pm0.0$ & P02 \\
B225 & 0.05& $  3$  & 13.87 & 12.89 & 0.15      & 4.65 & $-0.67\pm0.1$ & P02 \\
G001 & 0.20 &$162$  & 13.79 & \nodata &0.08     & 34.55& $-0.95\pm0.1$ & MEY \\
\multicolumn{9}{l}{STIS targets from program 8640}\\
B020 & 0.05 & $ 40$ & 14.97  & \nodata  & 0.08  &  7.37& $-1.07\pm0.1$ & HBK \\
B023 & 0.12 & $ 81$ & 14.12  & \nodata  & 0.36  &  4.45& $-0.92\pm0.1$ & HBK \\
B196 & 0.05 & $ 60$ & 17.36  & \nodata  & 0.24  & 8.10 & $-2.46\pm0.6$ & B00 \\
B236 & 0.16 & $149$ & 17.44  & \nodata  & 0.08  & 8.49 & $-1.20\pm0.6$ & AVE \\
B289 & 0.13 & $ 39$ & 16.27  & \nodata  & 0.08  & 16.88& $-1.20\pm0.6$ & AVE \\
B336 & 0.00 &\nodata& 17.91  & \nodata  & 0.08  & 12.91& $-1.20\pm0.6$ & AVE \\
B351 & 0.05 & $110$ & 17.61  & \nodata  & 0.11  & 12.57& $-1.43\pm0.2$ & B00 \\
B361 & 0.16 & $  0$ & 17.08  & \nodata  & 0.08  & 14.43& $-1.20\pm0.6$ & AVE \\
B363 & 0.15 & $173$ & 17.91  & \nodata  & 0.08  & 10.44& $-1.63\pm0.4$ & B00 \\
B373 & 0.09 & $ 22$ & 15.63  & \nodata  & 0.10  & 10.07& $-0.50\pm0.2$ & HBK \\
B396\tablenotemark{c}
     & 0.18 & $152$ & 17.45  & \nodata  & 0.08  & 17.27& $-1.20\pm0.6$ & AVE \\
B405 & 0.09 & $133$ & 15.25  & \nodata  & 0.12  & 18.20& $-1.80\pm0.3$ & HBK \\
B407 & 0.06 & $173$ & 16.11  & \nodata  & 0.16  & 19.76& $-0.85\pm0.3$ & HBK \\
B422 & 0.41 & $149$ & 17.98  & \nodata  & 0.08  & 16.34& $-1.20\pm0.6$ & AVE \\
B461 & 0.16 & $105$ & 17.44  & \nodata  & 0.47  & 10.87& $-1.36\pm0.5$ & B00 \\
B462 & 0.09 & $ 70$ & 18.03  & \nodata  & 0.08  &  3.74& $-1.20\pm0.6$ & AVE \\
G002 & 0.09 & $106$ & 15.93  & \nodata  & 0.08  & 33.62& $-1.70\pm0.4$ & HBK \\
G339 & 0.10 & $174$ & 17.35  & \nodata  & 0.08  & 28.68& $-1.20\pm0.6$ & AVE \\
G353 & 0.16 & $121$ & 17.31  & \nodata  & 0.08  & 26.32& $-1.20\pm0.6$ & AVE \\
\multicolumn{9}{l}{WFPC2-observed clusters from \citet{bhh02}}\\ 
B006 & 0.08  &  74  &  15.46 & \nodata & 0.13   &  6.39 & $-0.58\pm0.12$ & HBK \\  
B011 & 0.09  &  76  &  16.58 & \nodata & 0.08   &  7.68 & $-1.54\pm0.34$ & HBK \\  
B012 & 0.08  &  46  &  15.04 & \nodata & 0.11   &  5.74 & $-1.70\pm0.22$ & P02 \\  
B018 & 0.15  &   4  &  17.53 & \nodata & 0.08   &  9.27 & $-1.63\pm0.77$ & P02 \\  
B027 & 0.07  &  97  &  15.56 & \nodata & 0.16   &  5.99 & $-1.64\pm0.32$ & HBK \\  
B030 & 0.10  & 118  &  17.38 & \nodata & 0.66   &  5.64 & $-0.39\pm0.36$ & P02 \\  
B045 & 0.08  &  40  &  15.78 & \nodata & 0.18   &  4.87 & $-1.00\pm0.26$ & P02 \\  
B058 & 0.10  & 138  &  14.93 & \nodata & 0.12   &  6.94 & $-1.45\pm0.24$ & HBK \\  
B068 & 0.22  &  42  &  16.41 & \nodata & 0.45   &  4.31 & $-0.29\pm0.59$ & HBK \\  
B076 & 0.09  &  69  &  16.93 & \nodata & 0.21   &  2.85 & $-0.73\pm0.07$ & P02 \\  
B109 & 0.10  &  72  &  16.20 & \nodata & 0.08   &  1.39 & $-0.17\pm0.49$ & P02 \\  
B110 & 0.05  &  49  &  15.36 & \nodata & 0.16   &  2.92 & $-1.06\pm0.14$ & P02 \\  
B114 & 0.06  & 132  &  17.44 & \nodata & 0.16   &  0.88 & $-1.20\pm0.60$ & AVE \\  
B115 & 0.08  &  63  &  16.00 & \nodata & 0.08   &  0.64 & $-0.15\pm0.38$ & HBK \\  
B123 & 0.14  &  62  &  17.42 & \nodata & 0.08   &  1.28 & $-1.20\pm0.60$ & AVE \\  
B124 & 0.07  & 164  &  14.78 & \nodata & 0.08   &  0.21 & $-1.20\pm0.60$ & AVE \\  
B127 & 0.08  &  64  &  14.47 & \nodata & 0.15   &  0.32 & $-0.87\pm0.17$ & P02 \\  
B143 & 0.05  & 158  &  15.95 & \nodata & 0.13   &  0.97 & $ 0.09\pm0.42$ & HBK \\  
B155 & 0.12  &  80  &  18.01 & \nodata & 0.16   &  3.14 & $-1.20\pm0.60$ & AVE \\  
B156 & 0.05  &  67  &  16.97 & \nodata & 0.08   &  3.60 & $-1.51\pm0.38$ & P02 \\  
B160 & 0.18  &   2  &  18.08 & \nodata & 0.08   &  3.56 & $-1.17\pm1.25$ & P02 \\  
B231 & 0.17  & 136  &  17.25 & \nodata & 0.08   &  5.54 & $-1.49\pm0.41$ & P02 \\  
B232 & 0.18  &  42  &  15.65 & \nodata & 0.07   &  4.96 & $-1.81\pm0.17$ & P02 \\  
B233 & 0.11  &  74  &  15.72 & \nodata & 0.12   &  8.05 & $-1.59\pm0.32$ & HBK \\  
B234 & 0.07  &  71  &  16.78 & \nodata & 0.08   &  6.00 & $-0.95\pm0.13$ & P02 \\  
B240 & 0.16  &  98  &  15.18 & \nodata & 0.08   &  7.22 & $-1.76\pm0.18$ & HBK \\  
B264 & 0.26  & 142  &  17.58 & \nodata & 0.08   &  0.38 & $-1.20\pm0.60$ & AVE \\  
B268 & 0.12  & 103  &  18.31 & \nodata & 0.08   &  1.39 & $-1.20\pm0.60$ & AVE \\  
B279 & 0.20  &  79  &  18.55 & \nodata & 0.08   &  7.74 & $-1.20\pm0.60$ & AVE \\  
B311 & 0.09  &  54  &  15.45 & \nodata & 0.23   & 13.06 & $-1.93\pm0.08$ & P02 \\  
B315 & 0.13  & 159  &  16.47 & \nodata & 0.08   & 12.62 & $-2.11\pm0.53$ & HBK \\  
B317 & 0.11  &  66  &  16.57 & \nodata & 0.13   & 10.14 & $-2.12\pm0.36$ & HBK \\  
B319 & 0.0   &   0  &  17.61 & \nodata & 0.08   & 11.80 & $-2.27\pm0.47$ & B00 \\  
B324 & 0.0   &   0  &  18.45 & \nodata & 0.08   &  8.29 & $-1.20\pm0.60$ & AVE \\  
B328 & 0.27  & 159  &  17.86 & \nodata & 0.08   &  8.09 & $-1.22\pm0.80$ & HBK \\  
B330 & 0.14  & 102  &  17.72 & \nodata & 0.08   &  8.46 & $-1.20\pm0.60$ & AVE \\  
B331 & 0.24  &  70  &  18.19 & \nodata & 0.08   &  8.31 & $-1.44\pm0.69$ & B00 \\  
B333 & 0.23  &  26  &  18.84 & \nodata & 0.08   &  7.95 & $-1.20\pm0.60$ & AVE \\  
B338 & 0.06  & 102  &  14.20 & \nodata & 0.12   & 10.21 & $-1.38\pm0.09$ & HBK \\  
B343 & 0.09  &  70  &  16.31 & \nodata & 0.07   & 14.71 & $-1.49\pm0.17$ & HBK \\  
B358 & 0.12  &  63  &  15.22 & \nodata & 0.08   & 19.78 & $-1.83\pm0.22$ & HBK \\  
B368 & 0.0   &   0  &  17.92 & \nodata & 0.08   &  9.52 & $-1.20\pm0.60$ & AVE \\  
B374 & 0.21  & 106  &  18.31 & \nodata & 0.08   &  9.64 & $-1.90\pm0.67$ & P02 \\  
B379 & 0.09  &  55  &  16.18 & \nodata & 0.10   & 11.30 & $-0.70\pm0.35$ & HBK \\  
B384 & 0.20  & 121  &  15.75 & \nodata & 0.12   & 16.36 & $-0.66\pm0.22$ & HBK \\  
B386 & 0.08  & 140  &  15.55 & \nodata & 0.08   & 14.02 & $-1.57\pm0.17$ & P02 \\  
B468 & 0.0   &   0  &  17.79 & \nodata & 0.08   & 20.05 & $-1.20\pm0.60$ & AVE \\  
BH04 & 0.32  & 128  &  19.69 & \nodata & 0.08   &  8.33 & $-1.20\pm0.60$ & AVE \\  
BH05 & 0.19  & 123  &  16.06 & \nodata & 0.08   &  9.01 & $-1.20\pm0.60$ & AVE \\  
BH11 & 0     &   0  &  19.86 & \nodata & 0.08   &  9.41 & $-1.20\pm0.60$ & AVE \\ 
BH18 & 0.10  & 116  &  18.18 & \nodata & 0.08   & 12.35 & $-1.20\pm0.60$ & AVE \\ 
BH20 & 0.18  & 175  &  18.07 & \nodata & 0.08   &  4.44 & $-1.20\pm0.60$ & AVE \\ 
BH21 & 0.15  &  30  &  18.61 & \nodata & 0.08   &  5.01 & $-1.20\pm0.60$ & AVE \\ 
BH23 & 0.09  &  75  &  18.83 & \nodata & 0.08   &  1.28 & $-1.20\pm0.60$ & AVE \\ 
BH24 & 0.38  &  44  &  20.67 & \nodata & 0.08   &  5.37 & $-1.20\pm0.60$ & AVE \\ 
BH29 & 0     &  0   &  19.65 & \nodata & 0.08   &  8.71 & $-1.20\pm0.60$ & AVE \\ 
DAO38& 0.34  & 108  &  19.25 & \nodata & 0.08   &  9.43 & $-1.20\pm0.60$ & AVE \\  
M091 & 0     &   0  &  19.14 & \nodata & 0.08   &  9.31 & $-1.20\pm0.60$ & AVE \\  
NB39 & 0.13  &  28  &  17.94 & \nodata & 0.08   &  0.19 & $-1.20\pm0.60$ & AVE \\ 

\enddata
\tablenotetext{a}{F606W for ACS-observed clusters, except G001 (F555W).}
\tablenotetext{b}{HBK=\citet{hbk91}; P02=\citet{per02};
  B00=\citet{b00}; AVE=no measurement; ${\rm [Fe/H]}=-1.2\pm0.6$ assigned.} 
\tablenotetext{c}{These clusters appear close to the image edge; their measurements 
do not cover the full range of radial distance.}

\end{deluxetable}

\begin{deluxetable}{ccccccl}
\tablewidth{0pt}
\tablecaption{49 F606W, F814W, or V Intensity Profiles for 33 GCs in M31
\label{tab:M31sbprofs}}
\tablecolumns{7}
\tablehead{
\colhead{Name} & \colhead{Detector}  & \colhead{Filter} &
\colhead{$R$}  & \colhead{$I$} & \colhead{uncertainty} &
\colhead{Flag} \\
\colhead{}     & \colhead{}          & \colhead{}       & 
\colhead{[arcsec]}  & \colhead{[$L_\odot\ {\rm pc}^{-2}$]}  &
\colhead{[$L_\odot\ {\rm pc}^{-2}$]} & \colhead{}  \\
\colhead{(1)} & \colhead{(2)}  & \colhead{(3)}  & \colhead{(4)}  &
\colhead{(5)} & \colhead{(6)}  & \colhead{(7)}
}
\startdata
B056D & WFC & F814W &  0.0260 &  2233.332 & 101.584 &    OK  \\
B056D & WFC & F814W &  0.0287 &  2232.822 & 107.622 &   DEP  \\
B056D & WFC & F814W &  0.0315 &  2233.302 & 112.895 &   DEP  \\
B056D & WFC & F814W &  0.0347 &  2236.559 & 117.452 &   DEP  \\
B056D & WFC & F814W &  0.0381 &  2241.757 & 122.118 &   DEP  \\
B056D & WFC & F814W &  0.0420 &  2245.391 & 127.393 &   DEP  \\
B056D & WFC & F814W &  0.0461 &  2245.646 & 133.215 &   DEP  \\
B056D & WFC & F814W &  0.0508 &  2239.541 & 138.504 &   DEP  \\
B056D & WFC & F814W &  0.0558 &  2234.688 & 143.454 &    OK  \\
B056D & WFC & F814W &  0.0614 &  2229.836 & 148.906 &   DEP  \\
B056D & WFC & F814W &  0.0676 &  2227.591 & 154.365 &   DEP  \\
\enddata
\tablecomments{Table {\ref{tab:M31sbprofs}} is published in its entirety
in the electronic edition of the Journal. Only a small portion is shown here,
for guidance regarding its form and content. See text for description
of the FLAG column.}

\end{deluxetable}

\begin{deluxetable}{lllcclr}
\tablewidth{0pt}
\tablecaption{PSF Models:
  $I_{\rm PSF}=I_0\left[1+(R/r_0)^{\alpha}\right]^{-\beta/\alpha}$
\label{tab:psf}}
\tablecolumns{7}
\tablehead{
\colhead{Detector} & \colhead{Filter} 
& \colhead{$r_0$} & \colhead{$\alpha$} & \colhead{$\beta$} & \multicolumn{2}{c}{FWHM}\\
\colhead{} & \colhead{}  & \colhead{[arcsec]} & \colhead{} & \colhead{}  & 
\colhead{[arcsec]}  & \colhead{[px]}
}
\startdata
WFC & F606W & 0.0686 & 3 & 3.69 & $0.125$ &2.5   \\
WFC & F814W & 0.0783 & 3 & 3.56 & $0.145$ & 2.9   \\
\\
HRC & F555W & 0.0267 & 3 & 3.09 & $0.0527$& 2.1  \\
HRC & F606W & 0.0294 & 3 & 2.95 & $0.0593$& 2.4  \\
HRC & F814W & 0.0351 & 3 & 2.96 & $0.0706$& 2.8  \\
\\
STIS & CL     & 0.074& 2 & 4.2& $0.093$& 1.8 \\
\enddata
\end{deluxetable}

\begin{deluxetable}{lccccrrrrrrrr}
\tabletypesize{\scriptsize}
\rotate
\tablewidth{0pt}
\tablecaption{Basic Parameters of Fits to 49 Profiles of 33 GCs in M31
\label{tab:m31fits}}
\tablecolumns{13}
\tablehead{
\colhead{Name} & \colhead{Detector} & \colhead{Ext.} & \colhead{$V$ color} &
\colhead{$N_{\rm pts}$} & \colhead{Model} & \colhead{$\chi_{\rm min}^2$}   &
\colhead{$I_{\rm bkg}$} & \colhead{$W_0$} & \colhead{$c/n$}         &
\colhead{$\mu_0$} & \colhead{$\log\,r_0$} & \colhead{$\log\,r_0$}         \\
\colhead{} & \colhead{} & \colhead{[mag]} & \colhead{[mag]} & \colhead{}   &
\colhead{} & \colhead{} & \colhead{[$L_\odot\,{\rm pc}^{-2}$]} & \colhead{}  &
\colhead{} & \colhead{[mag arcsec$^{-2}$]} & \colhead{[arcsec]}              &
\colhead{[pc]}                                                            \\
\colhead{(1)}  & \colhead{(2)}  & \colhead{(3)}  & \colhead{(4)}  &
\colhead{(5)}  & \colhead{(6)}  & \colhead{(7)}  & \colhead{(8)}  &
\colhead{(9)}  & \colhead{(10)} & \colhead{(11)} & \colhead{(12)} &
\colhead{(13)}
}
\startdata
       B056D  & WFC/F606   & $0.224$  & $0.274\pm0.051$      & $52$       &
                K66  & $149.50$  & $20.38\pm0.76$  & $6.20^{+0.60}_{-0.60}$  &
                $1.31^{+0.16}_{-0.15}$  & $17.92^{+0.05}_{-0.06}$  &
                $-0.562^{+0.042}_{-0.042}$  & $0.016^{+0.042}_{-0.042}$ \\
          ~~        & ~~        & ~~        & ~~        & ~~         &
                W  & $131.14$  & $19.98\pm0.97$  & $6.10^{+0.70}_{-0.70}$  &
                $1.84^{+0.40}_{-0.28}$  & $17.95^{+0.05}_{-0.05}$  &
                $-0.526^{+0.044}_{-0.043}$  & $0.051^{+0.044}_{-0.043}$ \\
          ~~        & ~~        & ~~        & ~~        & ~~         &
                S  & $195.14$  & $20.92\pm0.60$       & ---  &
                $1.26^{+0.24}_{-0.20}$  & $17.45^{+0.23}_{-0.28}$  &
                $-0.896^{+0.161}_{-0.208}$  & $-0.318^{+0.161}_{-0.208}$ \\
       B056D  & WFC/F814   & $0.144$  & $1.286\pm0.064$      & $52$       &
                K66  & $175.21$  & $27.74\pm1.46$  & $5.80^{+0.80}_{-0.90}$  &
                $1.21^{+0.21}_{-0.20}$  & $16.88^{+0.08}_{-0.08}$  &
                $-0.561^{+0.068}_{-0.061}$  & $0.016^{+0.068}_{-0.061}$ \\
          ~~        & ~~        & ~~        & ~~        & ~~         &
                W  & $149.55$  & $27.25\pm1.55$  & $5.50^{+0.80}_{-1.10}$  &
                $1.59^{+0.35}_{-0.33}$  & $16.91^{+0.07}_{-0.06}$  &
                $-0.515^{+0.075}_{-0.053}$  & $0.063^{+0.075}_{-0.053}$ \\
          ~~        & ~~        & ~~        & ~~        & ~~         &
                S  & $224.33$  & $28.56\pm1.00$       & ---  &
                $1.14^{+0.28}_{-0.20}$  & $16.51^{+0.25}_{-0.35}$  &
                $-0.825^{+0.161}_{-0.243}$  & $-0.248^{+0.161}_{-0.243}$ \\
        B090  & WFC/F814   & $0.144$  & $1.114\pm0.064$      & $50$       &
                K66  & $105.21$  & $76.95\pm2.49$  & $7.50^{+0.60}_{-0.50}$  &
                $1.68^{+0.18}_{-0.15}$  & $15.71^{+0.09}_{-0.11}$  &
                $-0.898^{+0.045}_{-0.058}$  & $-0.320^{+0.045}_{-0.058}$ \\
          ~~        & ~~        & ~~        & ~~        & ~~         &
                W  & $107.84$  & $75.18\pm4.06$  & $7.50^{+0.70}_{-0.70}$  &
                $2.79^{+0.45}_{-0.55}$  & $15.76^{+0.09}_{-0.08}$  &
                $-0.860^{+0.052}_{-0.053}$  & $-0.282^{+0.052}_{-0.053}$ \\
          ~~        & ~~        & ~~        & ~~        & ~~         &
                S  & $124.24$  & $77.42\pm2.29$       & ---  &
                $2.30^{+0.60}_{-0.44}$  & $14.10^{+0.60}_{-0.81}$  &
                $-2.135^{+0.465}_{-0.684}$  & $-1.557^{+0.465}_{-0.684}$ \\
\enddata
\tablecomments{See text for column descriptions.
  Table \ref{tab:m31fits} is available in its entirety in the
  electronic 
  edition of the Journal. A short extract from it is shown here, for guidance
  regarding its form and content.}

\end{deluxetable}

\clearpage

\begin{deluxetable}{lcrrrrrrrrrr}
\tabletypesize{\scriptsize}
\rotate
\tablewidth{0pt}
\tablecaption{Derived Structural and Photometric Parameters of GCs in M31
\label{tab:m31phot}}
\tablecolumns{12}
\tablehead{
\colhead{Name} & \colhead{Detector} & \colhead{Model} &
\colhead{$\log\,r_{\rm tid}$} & \colhead{$\log\,R_c$} &
\colhead{$\log\,R_h$} & \colhead{$\log\,(R_h/R_c)$}   &
\colhead{$\log\,I_0$}  & \colhead{$\log\,j_0$}  & \colhead{$\log\,L_{V}$}  &
\colhead{$V_{\rm tot}$} & \colhead{$\log\,I_h$} \\
\colhead{} & \colhead{} & \colhead{} & \colhead{[pc]} & \colhead{[pc]} &
\colhead{[pc]} & \colhead{} &
\colhead{[$L_{\odot, V}\,{\rm pc}^{-2}$]}    &
\colhead{[$L_{\odot, V}\,{\rm pc}^{-3}$]}    &
\colhead{[$L_{\odot, V}$]} & \colhead{[mag]} &
\colhead{[$L_{\odot, V}\,{\rm pc}^{-2}$]}    \\
\colhead{(1)}  & \colhead{(2)}  & \colhead{(3)}  & \colhead{(4)}  &
\colhead{(5)}  & \colhead{(6)}  & \colhead{(7)}  & \colhead{(8)}  &
\colhead{(9)}  & \colhead{(10)} & \colhead{(11)} & \colhead{(12)}
}
\startdata
       B056D  & WFC/F606       & K66  & $1.32^{+0.12}_{-0.10}$  &
                $-0.016^{+0.032}_{-0.034}$  & $0.343^{+0.057}_{-0.038}$  &
                $0.358^{+0.092}_{-0.070}$  & $3.28^{+0.03}_{-0.03}$  &
                $2.99^{+0.06}_{-0.06}$  & $4.30^{+0.04}_{-0.03}$  &
                $18.54^{+0.08}_{-0.10}$  & $2.82^{+0.06}_{-0.08}$ \\
          ~~        & ~~         & W  & $1.89^{+0.36}_{-0.24}$  &
                $0.002^{+0.028}_{-0.030}$  & $0.383^{+0.102}_{-0.055}$  &
                $0.381^{+0.132}_{-0.083}$  & $3.27^{+0.03}_{-0.03}$  &
                $2.96^{+0.05}_{-0.05}$  & $4.34^{+0.06}_{-0.04}$  &
                $18.45^{+0.10}_{-0.16}$  & $2.77^{+0.08}_{-0.15}$ \\
          ~~        & ~~         & S  & $\infty$  &
                $-0.318^{+0.161}_{-0.208}$  & $0.313^{+0.040}_{-0.031}$  &
                $0.631^{+0.249}_{-0.192}$  & $3.47^{+0.11}_{-0.10}$  &
                $2.88^{+0.26}_{-0.20}$  & $4.26^{+0.03}_{-0.03}$  &
                $18.63^{+0.06}_{-0.07}$  & $2.84^{+0.05}_{-0.06}$ \\
       B056D  & WFC/F814       & K66  & $1.22^{+0.15}_{-0.13}$  &
                $-0.021^{+0.049}_{-0.049}$  & $0.288^{+0.058}_{-0.036}$  &
                $0.310^{+0.107}_{-0.085}$  & $3.29^{+0.04}_{-0.04}$  &
                $3.01^{+0.09}_{-0.09}$  & $4.25^{+0.04}_{-0.03}$  &
                $18.66^{+0.08}_{-0.10}$  & $2.88^{+0.06}_{-0.09}$ \\
          ~~        & ~~         & W  & $1.66^{+0.29}_{-0.25}$  &
                $0.000^{+0.042}_{-0.036}$  & $0.309^{+0.068}_{-0.045}$  &
                $0.309^{+0.104}_{-0.087}$  & $3.28^{+0.03}_{-0.04}$  &
                $2.97^{+0.07}_{-0.08}$  & $4.28^{+0.05}_{-0.04}$  &
                $18.59^{+0.10}_{-0.12}$  & $2.86^{+0.07}_{-0.10}$ \\
          ~~        & ~~         & S  & $\infty$  &
                $-0.248^{+0.161}_{-0.243}$  & $0.266^{+0.038}_{-0.022}$  &
                $0.514^{+0.281}_{-0.183}$  & $3.44^{+0.14}_{-0.10}$  &
                $2.82^{+0.31}_{-0.20}$  & $4.23^{+0.03}_{-0.03}$  &
                $18.72^{+0.07}_{-0.07}$  & $2.90^{+0.04}_{-0.07}$ \\
        B090  & WFC/F814       & K66  & $1.36^{+0.13}_{-0.11}$  &
                $-0.337^{+0.040}_{-0.053}$  & $0.259^{+0.113}_{-0.069}$  &
                $0.596^{+0.166}_{-0.109}$  & $3.83^{+0.05}_{-0.04}$  &
                $3.86^{+0.10}_{-0.08}$  & $4.40^{+0.06}_{-0.04}$  &
                $18.30^{+0.11}_{-0.15}$  & $3.08^{+0.10}_{-0.17}$ \\
          ~~        & ~~         & W  & $2.51^{+0.40}_{-0.50}$  &
                $-0.309^{+0.042}_{-0.045}$  & $0.445^{+0.329}_{-0.198}$  &
                $0.754^{+0.374}_{-0.240}$  & $3.81^{+0.04}_{-0.04}$  &
                $3.82^{+0.08}_{-0.08}$  & $4.51^{+0.15}_{-0.10}$  &
                $18.02^{+0.26}_{-0.36}$  & $2.82^{+0.30}_{-0.51}$ \\
          ~~        & ~~         & S  & $\infty$  &
                $-1.557^{+0.465}_{-0.684}$  & $0.259^{+0.101}_{-0.068}$  &
                $1.816^{+0.785}_{-0.534}$  & $4.47^{+0.33}_{-0.24}$  &
                $4.90^{+0.92}_{-0.62}$  & $4.38^{+0.06}_{-0.05}$  &
                $18.35^{+0.12}_{-0.15}$  & $3.06^{+0.10}_{-0.15}$ \\
\enddata
\tablecomments{See text for column descriptions. 
  Table \ref{tab:m31phot} is available in its entirety in the
  electronic 
  edition of the Journal. A short extract from it is shown here, for guidance
  regarding its form and content.}

\end{deluxetable}

\begin{deluxetable}{lccrrrrrrrrrr}
\tabletypesize{\scriptsize}
\rotate
\tablewidth{0pt}
\tablecaption{Derived Dynamical Parameters of GCs in M31
\label{tab:m31mass}}
\tablecolumns{13}
\tablehead{
\colhead{Name} & \colhead{Detector} & \colhead{$\Upsilon_V^{\rm pop}$}     &
\colhead{Model} & \colhead{$\log\,M_{\rm tot}$} & \colhead{$\log\,E_b$}  &
\colhead{$\log\,\Sigma_0$} & \colhead{$\log\,\rho_0$} & 
\colhead{$\log\,\Sigma_{\rm h}$} & \colhead{$\log\,\sigma_{{\rm p},0}$}     &
\colhead{$\log\,v_{{\rm esc},0}$} & \colhead{$\log\,t_{\rm rh}$} &
\colhead{$\log\,f_0$}                                            \\
\colhead{} & \colhead{} & \colhead{[$M_\odot\,L_{\odot,V}^{-1}$]}  &
\colhead{} & \colhead{[$M_\odot$]} & \colhead{[erg]} &
\colhead{[$M_\odot$ pc$^{-2}$]} & \colhead{[$M_\odot$ pc$^{-3}$]} &
\colhead{[$M_\odot$ pc$^{-2}$]} & \colhead{[km~s$^{-1}$]} & \colhead{[km~s$^{-1}$]} &
\colhead{[yr]} & \colhead{[$M_\odot$ (pc km~s$^{-1}$)$^{-3}$]}                                      \\
\colhead{(1)}  & \colhead{(2)}  & \colhead{(3)}  & \colhead{(4)}  &
\colhead{(5)}  & \colhead{(6)}  & \colhead{(7)}  & \colhead{(8)}  &
\colhead{(9)}  & \colhead{(10)} & \colhead{(11)}  & \colhead{(12)} & \colhead{(13)}
}
\startdata
       B056D  & WFC/F606  & $2.005^{+0.617}_{-0.263}$       & K66  &
                $4.60^{+0.12}_{-0.07}$  & $49.27^{+0.17}_{-0.09}$  &
                $3.58^{+0.12}_{-0.07}$  & $3.29^{+0.13}_{-0.08}$  &
                $3.12^{+0.13}_{-0.10}$  & $0.526^{+0.059}_{-0.033}$  &
                $1.101^{+0.059}_{-0.033}$  & $8.43^{+0.12}_{-0.07}$  &
                $0.484^{+0.103}_{-0.087}$ \\
          ~~        & ~~        & ~~         & W  &
                $4.64^{+0.13}_{-0.07}$  & $49.29^{+0.17}_{-0.09}$  &
                $3.57^{+0.12}_{-0.07}$  & $3.26^{+0.13}_{-0.08}$  &
                $3.07^{+0.14}_{-0.16}$  & $0.530^{+0.059}_{-0.033}$  &
                $1.116^{+0.059}_{-0.033}$  & $8.51^{+0.19}_{-0.10}$  &
                $0.434^{+0.097}_{-0.078}$ \\
          ~~        & ~~        & ~~         & S  &
                $4.56^{+0.12}_{-0.07}$  & $49.24^{+0.17}_{-0.09}$  &
                $3.77^{+0.16}_{-0.11}$  & $3.18^{+0.27}_{-0.20}$  &
                $3.14^{+0.13}_{-0.09}$  & $0.518^{+0.060}_{-0.036}$  &
                $1.123^{+0.059}_{-0.032}$  & $8.37^{+0.09}_{-0.06}$  &
                $0.433^{+0.377}_{-0.267}$ \\
       B056D  & WFC/F814  & $2.005^{+0.617}_{-0.263}$       & K66  &
                $4.56^{+0.12}_{-0.07}$  & $49.24^{+0.17}_{-0.09}$  &
                $3.60^{+0.12}_{-0.07}$  & $3.31^{+0.15}_{-0.11}$  &
                $3.18^{+0.13}_{-0.11}$  & $0.529^{+0.060}_{-0.034}$  &
                $1.095^{+0.060}_{-0.035}$  & $8.31^{+0.12}_{-0.07}$  &
                $0.485^{+0.134}_{-0.127}$ \\
          ~~        & ~~        & ~~         & W  &
                $4.58^{+0.13}_{-0.07}$  & $49.25^{+0.17}_{-0.09}$  &
                $3.58^{+0.12}_{-0.07}$  & $3.27^{+0.13}_{-0.10}$  &
                $3.17^{+0.13}_{-0.12}$  & $0.534^{+0.060}_{-0.034}$  &
                $1.110^{+0.060}_{-0.034}$  & $8.35^{+0.13}_{-0.09}$  &
                $0.423^{+0.108}_{-0.113}$ \\
          ~~        & ~~        & ~~         & S  &
                $4.53^{+0.12}_{-0.07}$  & $49.22^{+0.17}_{-0.09}$  &
                $3.74^{+0.18}_{-0.12}$  & $3.12^{+0.31}_{-0.20}$  &
                $3.20^{+0.12}_{-0.09}$  & $0.524^{+0.060}_{-0.037}$  &
                $1.121^{+0.060}_{-0.034}$  & $8.27^{+0.09}_{-0.05}$  &
                $0.324^{+0.425}_{-0.252}$ \\
        B090  & WFC/F814  & $2.005^{+0.617}_{-0.263}$       & K66  &
                $4.70^{+0.13}_{-0.08}$  & $49.56^{+0.17}_{-0.09}$  &
                $4.13^{+0.13}_{-0.07}$  & $4.16^{+0.15}_{-0.10}$  &
                $3.38^{+0.16}_{-0.18}$  & $0.638^{+0.060}_{-0.034}$  &
                $1.240^{+0.060}_{-0.034}$  & $8.33^{+0.20}_{-0.12}$  &
                $1.031^{+0.132}_{-0.096}$ \\
          ~~        & ~~        & ~~         & W  &
                $4.81^{+0.19}_{-0.12}$  & $49.59^{+0.17}_{-0.10}$  &
                $4.12^{+0.12}_{-0.07}$  & $4.12^{+0.14}_{-0.10}$  &
                $3.12^{+0.32}_{-0.52}$  & $0.643^{+0.060}_{-0.034}$  &
                $1.254^{+0.060}_{-0.034}$  & $8.65^{+0.56}_{-0.34}$  &
                $0.962^{+0.121}_{-0.104}$ \\
          ~~        & ~~        & ~~         & S  &
                $4.68^{+0.13}_{-0.08}$  & $49.52^{+0.17}_{-0.09}$  &
                $4.78^{+0.35}_{-0.25}$  & $5.20^{+0.92}_{-0.63}$  &
                $3.36^{+0.15}_{-0.16}$  & $0.551^{+0.074}_{-0.078}$  &
                $1.274^{+0.062}_{-0.036}$  & $8.32^{+0.18}_{-0.12}$  &
                $2.707^{+1.425}_{-0.929}$ \\
\enddata
\tablecomments{See text for column descriptions.
  Table \ref{tab:m31mass} is available in its entirety in the
  electronic 
  edition of the Journal. A short extract from it is shown here, for guidance
  regarding its form and content.}

\end{deluxetable}

\begin{deluxetable}{lcrrrrr}
\tabletypesize{\scriptsize}
\tablewidth{0pt}
\tablecaption{Galactocentric Radii and $\kappa$-Space Parameters for GCs in M31
\label{tab:m31kappa}}
\tablecolumns{7}
\tablehead{
\colhead{Name} & \colhead{Detector} & \colhead{$R_{\rm gc}$}        &
\colhead{Model} &
\colhead{$\kappa_{m,1}$} & \colhead{$\kappa_{m,2}$} & \colhead{$\kappa_{m,3}$}  \\
\colhead{} & \colhead{} & \colhead{[kpc]} & \colhead{} & \colhead{} &
\colhead{} \\
\colhead{(1)}  & \colhead{(2)}  & \colhead{(3)}  & \colhead{(4)}  &
\colhead{(5)}  & \colhead{(6)}  & \colhead{(7)}
}
\startdata
       B056D  & WFC/F606      & 4.20       & K66  &
                $-1.136^{+0.089}_{-0.049}$  & $4.059^{+0.158}_{-0.124}$  &
                $0.341^{+0.007}_{-0.002}$ \\
          ~~        & ~~        & ~~         & W  &
                $-1.101^{+0.105}_{-0.055}$  & $4.011^{+0.169}_{-0.183}$  &
                $0.347^{+0.017}_{-0.005}$ \\
          ~~        & ~~        & ~~         & S  &
                $-1.168^{+0.084}_{-0.047}$  & $4.085^{+0.156}_{-0.112}$  &
                $0.336^{+0.001}_{-0.007}$ \\
       B056D  & WFC/F814      & 4.20       & K66  &
                $-1.169^{+0.089}_{-0.049}$  & $4.136^{+0.160}_{-0.130}$  &
                $0.340^{+0.005}_{-0.000}$ \\
          ~~        & ~~        & ~~         & W  &
                $-1.147^{+0.093}_{-0.051}$  & $4.120^{+0.164}_{-0.138}$  &
                $0.343^{+0.008}_{-0.001}$ \\
          ~~        & ~~        & ~~         & S  &
                $-1.192^{+0.084}_{-0.048}$  & $4.156^{+0.152}_{-0.113}$  &
                $0.337^{+0.000}_{-0.005}$ \\
        B090  & WFC/F814      & 3.15       & K66  &
                $-1.036^{+0.111}_{-0.064}$  & $4.401^{+0.185}_{-0.207}$  &
                $0.366^{+0.027}_{-0.014}$ \\
          ~~        & ~~        & ~~         & W  &
                $-0.898^{+0.239}_{-0.141}$  & $4.115^{+0.358}_{-0.565}$  &
                $0.415^{+0.100}_{-0.051}$ \\
          ~~        & ~~        & ~~         & S  &
                $-1.159^{+0.085}_{-0.055}$  & $4.315^{+0.205}_{-0.235}$  &
                $0.277^{+0.033}_{-0.054}$ \\
\enddata
\tablecomments{See text for column descriptions, particularly difference
between $\kappa$ definitions used here and those of \citet{bbf92}.
Table \ref{tab:m31kappa} is available in its entirety in the
electronic edition of the Journal. A short extract from it is shown here,
for guidance regarding its form and content.}
\end{deluxetable}

\begin{deluxetable}{lcrrrrrrrrr}
\tabletypesize{\scriptsize}
\rotate
\tablewidth{0pt}
\tablecaption{Predicted Aperture Velocity Dispersions for GCs in M31
\label{tab:m31vels}}
\tablecolumns{11}
\tablehead{
\colhead{Name} & \colhead{Detector} & \colhead{$\Upsilon_V^{\rm pop}$}     &
\colhead{Model} & \colhead{$\log\,R_h$} & \colhead{$\log\,R_h$}  &
\colhead{$\log\,\sigma_{\rm ap}(R_h/8)$} &
\colhead{$\log\,\sigma_{\rm ap}(R_h/4)$} &
\colhead{$\log\,\sigma_{\rm ap}(R_h)$} &
\colhead{$\log\,\sigma_{\rm ap}(4\,R_h)$} &
\colhead{$\log\,\sigma_{\rm ap}(8\,R_h)$} \\
\colhead{} & \colhead{} & \colhead{[$M_\odot\,L_{\odot,V}^{-1}$]}  &
\colhead{} & \colhead{[pc]} & \colhead{[arcsec]}                   &
\colhead{[km~s$^{-1}$]} &
\colhead{[km~s$^{-1}$]} & \colhead{[km~s$^{-1}$]} &
\colhead{[km~s$^{-1}$]} & \colhead{[km~s$^{-1}$]} \\
\colhead{(1)}  & \colhead{(2)}  & \colhead{(3)}  & \colhead{(4)}  &
\colhead{(5)}  & \colhead{(6)}  & \colhead{(7)}  & \colhead{(8)}  &
\colhead{(9)}  & \colhead{(10)} & \colhead{(11)}
}
\startdata
       B056D  & WFC/F606  & $2.005^{+0.617}_{-0.263}$       & K66     &
                0.343  & $-0.235^{+0.057}_{-0.038}$  &
                $0.525^{+0.059}_{-0.033}$  & $0.523^{+0.059}_{-0.033}$  &
                $0.501^{+0.059}_{-0.033}$  & $0.454^{+0.060}_{-0.033}$  &
                $0.446^{+0.060}_{-0.034}$ \\
          ~~        & ~~        & ~~         & W     &
                0.383  & $-0.194^{+0.102}_{-0.055}$  &
                $0.529^{+0.059}_{-0.033}$  & $0.526^{+0.059}_{-0.033}$  &
                $0.499^{+0.060}_{-0.034}$  & $0.453^{+0.060}_{-0.034}$  &
                $0.441^{+0.060}_{-0.036}$ \\
          ~~        & ~~        & ~~         & S     &
                0.313  & $-0.265^{+0.040}_{-0.031}$  &
                $0.494^{+0.060}_{-0.033}$  & $0.508^{+0.060}_{-0.033}$  &
                $0.503^{+0.060}_{-0.033}$  & $0.454^{+0.060}_{-0.034}$  &
                $0.450^{+0.060}_{-0.034}$ \\
       B056D  & WFC/F814  & $2.005^{+0.617}_{-0.263}$       & K66     &
                0.288  & $-0.289^{+0.058}_{-0.036}$  &
                $0.528^{+0.060}_{-0.034}$  & $0.527^{+0.060}_{-0.035}$  &
                $0.505^{+0.060}_{-0.035}$  & $0.458^{+0.060}_{-0.035}$  &
                $0.452^{+0.061}_{-0.036}$ \\
          ~~        & ~~        & ~~         & W     &
                0.309  & $-0.269^{+0.068}_{-0.045}$  &
                $0.533^{+0.060}_{-0.034}$  & $0.531^{+0.060}_{-0.034}$  &
                $0.506^{+0.060}_{-0.035}$  & $0.460^{+0.060}_{-0.035}$  &
                $0.450^{+0.061}_{-0.036}$ \\
          ~~        & ~~        & ~~         & S     &
                0.266  & $-0.311^{+0.038}_{-0.022}$  &
                $0.501^{+0.061}_{-0.036}$  & $0.514^{+0.061}_{-0.035}$  &
                $0.508^{+0.060}_{-0.035}$  & $0.459^{+0.060}_{-0.035}$  &
                $0.456^{+0.060}_{-0.036}$ \\
        B090  & WFC/F814  & $2.005^{+0.617}_{-0.263}$       & K66     &
                0.259  & $-0.318^{+0.113}_{-0.069}$  &
                $0.636^{+0.060}_{-0.034}$  & $0.633^{+0.060}_{-0.034}$  &
                $0.608^{+0.060}_{-0.035}$  & $0.556^{+0.060}_{-0.037}$  &
                $0.542^{+0.061}_{-0.038}$ \\
          ~~        & ~~        & ~~         & W     &
                0.445  & $-0.132^{+0.329}_{-0.198}$  &
                $0.639^{+0.060}_{-0.036}$  & $0.632^{+0.060}_{-0.039}$  &
                $0.595^{+0.062}_{-0.047}$  & $0.543^{+0.063}_{-0.053}$  &
                $0.524^{+0.065}_{-0.055}$ \\
          ~~        & ~~        & ~~         & S     &
                0.259  & $-0.318^{+0.101}_{-0.068}$  &
                $0.608^{+0.060}_{-0.033}$  & $0.621^{+0.060}_{-0.033}$  &
                $0.601^{+0.060}_{-0.035}$  & $0.548^{+0.060}_{-0.036}$  &
                $0.537^{+0.061}_{-0.037}$ \\
\enddata
\tablecomments{See text for column descriptions. Table \ref{tab:m31vels} is 
  available in its entirety in the
  electronic edition of the Journal. A short extract from it is shown here,
  for guidance regarding its form and content.}

\end{deluxetable}

\begin{deluxetable}{lrrl}
\tabletypesize{\scriptsize}
\tablewidth{0pt}
\tablecaption{Measured Aperture Velocity Dispersions for GCs in M31
\label{tab:m31veldisp}}
\tablecolumns{4}
\tablehead{
\colhead{Name} & \colhead{$\sigma_{\rm ap}$} &
\colhead{Aperture} & \colhead{Source}\\
\colhead{}  & \colhead{[km~s$^{-1}$]} &
\colhead{[arcsec]} & \colhead{}
}
\startdata
\sidehead{Clusters in the present sample}\\
B006    &    $11.56\pm0.11$&  1.60 &   \citet{djo97}\\ 	 
B006    &    $10.6 \pm0.4 $&  1.80 &   \citet{dg97}\\ 	 
B012    &    $16.15\pm0.45$&  1.60 &   \citet{djo97}\\ 	 
B020    &    $14.27\pm0.24$&  1.60 &   \citet{djo97}\\ 	 
B020    &    $15.3 \pm0.5 $&  1.55 &   \citet{dg97}\\ 	 
B020    &    $18   \pm3   $&  1.2  &   \citet{pet89}\\ 	 
B023    &    $25.46\pm0.90$&  1.60 &   \citet{djo97}\\ 	 
B023    &    $24   \pm6   $&  1.2  &   \citet{pet89}\\ 	 
B037    &    $19.55\pm3.50$&  2.77 &   \citet{cohen06}\\ 
B045    &    $ 9.82\pm0.18$&  1.60 &   \citet{djo97}\\ 	 
B045    &    $ 8.7 \pm0.5 $&  1.80 &   \citet{dg97}\\ 	 
B082    &    $25   \pm5   $&  1.2  &   \citet{pet89}\\ 	 
B147    &    $11   \pm2   $&  1.2  &   \citet{pet89}\\ 	 
B158    &    $20.50\pm0.26$&  1.60 &   \citet{djo97}\\ 	 
B158    &    $21.9 \pm1.3 $&  1.55 &   \citet{dg97}\\ 	 
B158    &    $40   \pm6   $&  1.2  &   \citet{pet89}\\ 	 
B225    &    $25.94\pm0.38$&  1.60 &   \citet{djo97}\\ 	 
B225    &    $26.9 \pm0.5 $&  1.80 &   \citet{dg97}\\ 	 
B225    &    $<40      $&  1.2  &   \citet{pet89}\\ 	 
B240    &    $11.92\pm0.22$&  1.60 &   \citet{djo97}\\ 	 
B289    &    $ 8.43\pm0.44$&  1.60 &   \citet{djo97}\\ 	 
B343    &    $ 9.08\pm0.39$&  1.60 &   \citet{djo97}\\ 	 
B343    &    $10.2 \pm1.7 $&  1.55 &   \citet{dg97}\\ 	 
B358    &    $ 8.11\pm0.36$&  1.60 &   \citet{djo97}\\ 	 
B358    &    $ 7.1 \pm1.8 $&  1.55 &   \citet{dg97}\\ 	 
B373    &    $12.58\pm0.15$&  1.60 &   \citet{djo97}\\ 	 
B379    &    $ 8.15\pm0.20$&  1.60 &   \citet{djo97}\\ 	 
B384    &    $10.10\pm0.16$&  1.60 &   \citet{djo97}\\ 	 
B384    &    $ 9.1 \pm0.5 $&  1.55 &   \citet{dg97}\\ 	 
B386    &    $11.49\pm0.24$&  1.60 &   \citet{djo97}\\ 	 
B405    &    $ 8.57\pm0.45$&  1.60 &   \citet{djo97}\\ 	 
B407    &    $ 9.52\pm0.11$&  1.60 &   \citet{djo97}\\ 	 
G001    &    $25.06\pm0.32$&  1.60 &   \citet{djo97}\\   
G001    &    $21.47\pm1.60$&  2.77 &   \citet{cohen06}\\ 
G002    &    $ 9.70\pm0.29$&  1.60 &   \citet{djo97}\\ 	 
\sidehead{Clusters not in the present sample}		 
B019    &    $19   \pm3   $&  1.2  &   \citet{pet89}\\   
B163    &    $21   \pm4   $&  1.2  &   \citet{pet89}\\   
B171    &    $18   \pm3   $&  1.2  &   \citet{pet89}\\   
B193    &    $13.20\pm0.24$&  1.60 &   \citet{djo97}\\   
B193    &    $12   \pm3   $&  1.2  &   \citet{pet89}\\   
B218    &    $17.62\pm0.24$&  1.60 &   \citet{djo97}\\   
B218    &    $16.3 \pm0.8 $&  1.55 &   \citet{dg97}\\  	 
B224    &    $<10 $&  1.2  &   \citet{pet89}\\   
B381    &    $<10 $&  1.2  &   \citet{pet89}\\   
\enddata

\end{deluxetable}

\begin{deluxetable}{lllrlllrl}
\tablewidth{0pt}
\tablecaption{Fundamental plane fits
\label{tab:fp_fits}}
\tablecolumns{9}
\rotate
\tablehead{
\colhead{Galaxy} & 
\multicolumn{3}{c}{$\log R_h = C+ \gamma \log(R_{\rm gc}^*)$} & 
\multicolumn{3}{c}{$\log E_b^{*} = a(\log M)+b$} & \colhead{} & \colhead{}\\
\colhead{} & \colhead{$C$} & \colhead{$\gamma$} & \colhead{RMS} &
\colhead{$a$} & \colhead{$b$} & \colhead{RMS} & \colhead{$\delta \log E_b^{*}$} & \colhead{$N$}
}
\startdata
Milky Way     &$0.57\pm0.02$ & $0.47\pm0.05$ & 0.21 &  $2.04\pm0.04$ & $39.63\pm0.21$ & 0.21& $0.02\pm0.02$  &85\\ 
NGC~5128      &$0.58\pm0.02$ & $0.19\pm0.07$ & 0.19 &  $2.09\pm0.04$ & $39.36\pm0.20$ & 0.19& $-0.01\pm0.02$  &104\\ 
M31	      &$0.43\pm0.02$ & $0.20\pm0.04$ & 0.17 &  $1.92\pm0.03$ & $40.43\pm0.15$ & 0.17& $0.15\pm0.02$  &84\\ 
M31(ACS/STIS) &$0.47\pm0.02$ & $0.11\pm0.04$ & 0.09 &  $1.96\pm0.03$ & $40.19\pm0.15$ & 0.10& $0.09\pm0.02$  &33\\ 
MCs+Fornax    &$0.65\pm0.04$ & $0.41\pm0.10$ & 0.17 &  $2.25\pm0.12$ & $38.48\pm0.65$ & 0.15& $-0.07\pm0.04$  &18\\ 
\enddata
\end{deluxetable}

\end{document}